\documentclass[
pra,superscriptaddress,reprint
]{revtex4-2}

\usepackage{graphicx}
\usepackage{hyperref}

\usepackage{amsmath,amssymb,amsfonts}
\usepackage{bm}
\usepackage{bbold} 
\usepackage{ulem}

\makeatletter
\@fleqnfalse\makeatother
\usepackage{dcolumn}
\usepackage{xcolor}
\usepackage{placeins}

\renewcommand{\vec}[1]{{\mathbf#1}}
\newcommand{\braket}[1]{\left\langle#1\right\rangle}

\usepackage{bibunits}
\defaultbibliography{bibliography}
\defaultbibliographystyle{apsrev4-2}

\begin{document}

\title{Delocalization transition for light in two dimensions}

\author{S. Lucas}%
\affiliation{Universit\'{e} Grenoble Alpes, CNRS, LPMMC, 38000 Grenoble, France}

\author{C. Miniatura}
\affiliation{Universit\'{e} C\^{o}te d'Azur, CNRS, INPHYNI, Nice, France}

\author{S.E. Skipetrov}%
\affiliation{Universit\'{e} Grenoble Alpes, CNRS, LPMMC, 38000 Grenoble, France}

\date{\today}

\begin{abstract}

Common belief, confirmed by existing experiments, is that arbitrarily weak disorder should lead to spatial localization of eigenmodes of scalar wave equations when wave propagation is two-dimensional (2D). We predict that, contrary to this belief, a localization-delocalization transition can take place for light scattered by two-level atoms placed at random positions in the middle plane of a parallel-plate 2D waveguide fed by its fundamental transverse-magnetic (TM) mode (electric field polarized perpendicular to the waveguide and to the atomic plane). This transition, driven by near-field dipole-dipole interactions between atoms, occurs upon increasing the areal number density of atoms beyond some critical value. A finite-size scaling analysis of the transition yields an estimate of its critical exponent $\nu=1.4\pm0.2$.

\end{abstract}

\maketitle

\begin{bibunit}

\quad Anderson localization---suppression of wave transport by destructive interferences in disordered media---was first discovered for electrons in disordered solids~\cite{AbsDiffCertain}, and has since been shown to play an important role across many areas of physics~\cite{lagendijk09,abrahams10}. For electromagnetic waves in general and light in particular, Anderson localization has been robustly demonstrated in quasi-one-dimensional disordered waveguides~\cite{chabanov00,wang11} as well as in two-dimensional (2D) geometries~\cite{dalichaouch91,schwartz07,laurent07,riboli11}, where the scaling theory predicts all states to be spatially localized in the limit of infinite sample size and in the absence of any discrete symmetry (time-reversal, particle-hole or chiral) breaking~\cite{Gang4}. In the most interesting case of three-dimensional (3D) disorder for which a transition from extended to localized states is expected~\cite{evers08}, Anderson localization of light remains elusive due to the vectorial nature of light and the proliferation of longitudinal optical modes favored by strong disorder~\cite{skipetrov_absence_2014,LongiModes_bart_sergey, yamilov_anderson_2023}. Disordered photonic crystals~\cite{john87,skipetrov20}, ensembles of cold atoms in strong external magnetic fields~\cite{skipetrov_magnetic-field-driven_2015,skipetrov_localization_2018}, or disordered materials made of highly conducting, metallic particles~\cite{yamilov_anderson_2023,AndersonTrans3D} have been theoretically proposed to overcome this complication. Longitudinal optical modes and the associated enhanced near-field coupling between scatterers have been demonstrated to prevent Anderson localization even in 2D, for in-plane polarization of light~\cite{maximo_spatial_2015}. These results suggest that in disordered media, light behaves fundamentally differently from other waves (matter waves, sound, elastic waves, \textit{etc}.).

\begin{figure}
    \centering
    \includegraphics[width=1\columnwidth]{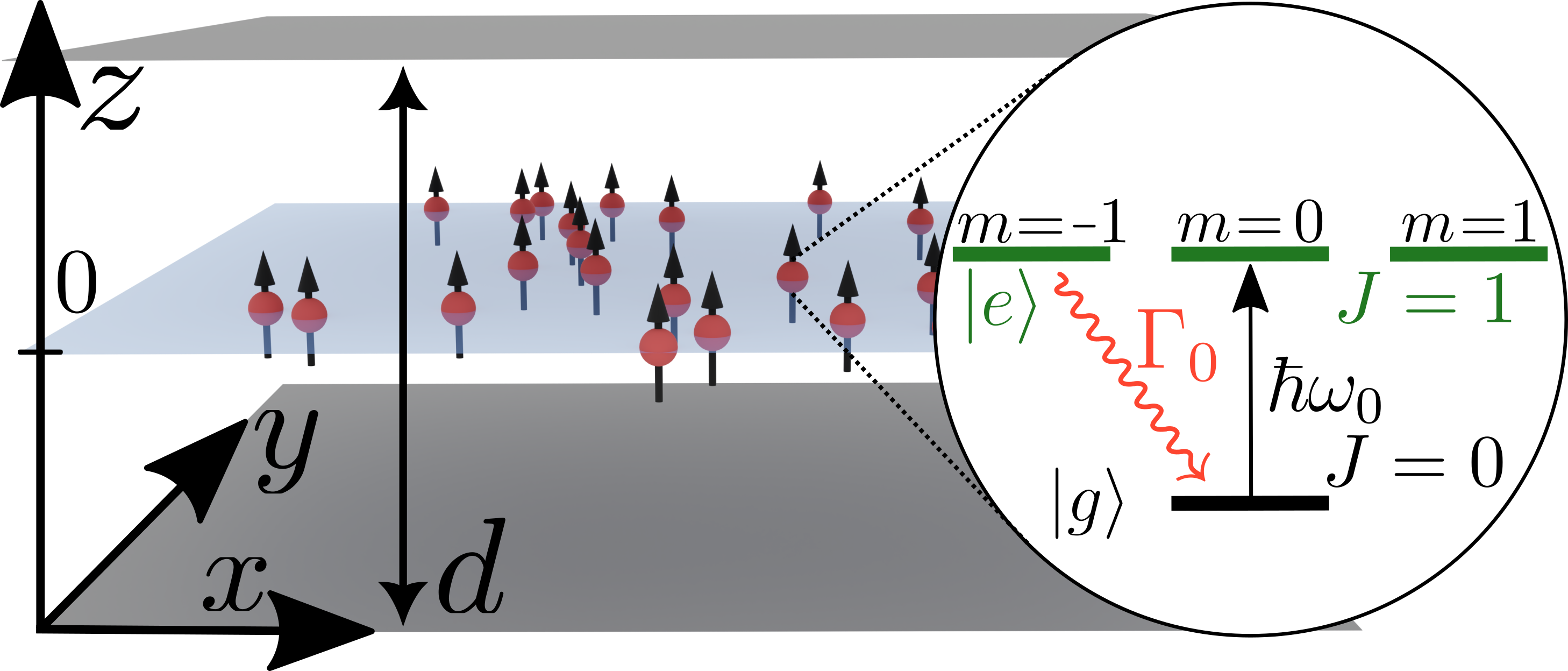}
    \caption{$N$ two-level atoms (red dots) are randomly distributed in the $z=0$ plane in the middle of a planar waveguide formed by two perfectly reflecting parallel mirrors (gray plates) separated by a distance $d < \pi/k_0$. Such a planar waveguide only admits propagating TM$_{0}$ modes whose electric field, perpendicular to the mirrors, polarizes the atomic dipoles along $z$ axis (black arrows) and is subsequently multiply scattered by the atoms.
    }\label{fig:geometry}
\end{figure}

\quad In this Letter, we show that in low-dimensional systems, the fundamental difference between light and other types of waves can have even more profound consequences than in 3D.
We consider collective excitations of a disordered ensemble of point-like resonant scatterers (atoms) confined between two closely spaced, parallel perfectly reflecting plates forming a planar 2D waveguide (see Fig.~\ref{fig:geometry}). Focusing on excitations polarized perpendicular to the atomic plane (the so-called $s$ polarization \cite{maximo_spatial_2015}), we discover a localization-delocalization \textit{transition} upon increasing the areal number density of atoms $\rho$.
Due to the strong light-atom coupling, delocalization of atomic excitations implies delocalization of light propagating in the waveguide
(see Sec.\ S10 of Supplementary Material \cite{sm}). Provided that atoms are in the middle plane of the waveguide and that the mirror spacing $d < \pi/k_0$, where $k_0$ is the optical wavenumber at the single-atom resonance in the free space, $s$-polarized atomic excitations couple only to transverse magnetic (TM) modes of the empty waveguide, of which only TM$_{0}$ are propagating modes with real wavevectors. Thus, this system is the closest realization of a 2D geometry achievable with atomic scatterers that are intrinsically 3D.
Its direction being fixed, the atomic polarization is described by a scalar 2D field subject to Anderson localization for low $\rho$, as predicted for scalar waves in 2D by the scaling theory~\cite{Gang4}. Surprisingly, however, increasing $\rho$ has a counter-intuitive effect of suppressing Anderson localization and inducing a delocalization transition that is due to the increased coupling between atoms and takes place at a critical density $\rho_c \sim k_0^2$.
Reaching such densities in a planar atomic sample confined in a 2D waveguide is experimentally demanding but not impossible: cold atoms have been confined to thin slabs in the free space at densities up to $\rho/k_0^2 \sim 2$, comparable to $\rho_c$ ~\protect\cite{vatre26,res_light_diff_disordered_layer}.
Planar waveguides satisfying $k_0 d < \pi$ are also within reach of current nanophotonics platforms~\protect\cite{Alaeian2020}.
Although the field of atomic polarization remains scalar and strictly 2D, the 3D dipole-dipole near-field coupling between the atoms violates the assumptions of the scalar 2D scaling theory~\cite{Gang4}, hinting at a critical behavior possibly escaping the standard ten-fold Altland-Zirnbauer classification~\cite{zirnbauer96,altland97,evers08}.

\quad We consider an ensemble of $N$ identical two-level atoms featuring an allowed electric-dipole transition between a non-degenerate ground state $|g_a\rangle$ with an angular momentum $J_g=0$ and a three-fold degenerate excited state $|e_{a m}\rangle$ with $J_e=1$, $a = 1,\ldots,N$, $m = 0, \pm 1$, and the quantization axis chosen along $z$. In terms of raising and lowering operators $\sigma_{a m}^+ = |e_{a m}\rangle\langle g_a|$ and $\sigma_{a m}^- = {(\sigma_{a m}^+)}^\dagger = |g_a\rangle\langle e_{a m}|$, the atomic Hamiltonian is $H_0 = \hbar\omega_0 \sum_{a, m} |e_{a m}\rangle \langle e_{a m}| = \hbar\omega_0 \sum_{a, m} \sigma_{a m}^+\sigma_{a m}^-$, where $\omega_0$ is the resonant frequency of the transition. Note that local atomic operators are implicitly extended to the full many-body Hilbert space by tensoring with the identity on all other atoms. Atoms are placed at random, uncorrelated positions $\{\vec{r}_a \}$ within a disk of radius $R$ in the plane $z=0$ in the middle of a thin planar waveguide (see Fig.~\ref{fig:geometry}), which ensures that in-plane ($p$-polarized) and out-of-plane ($s$-polarized) collective atomic excitations remain decoupled. Accounting for the coupling between the atoms and the electromagnetic field within the Markov approximation, restricting consideration to $s$ polarization, and tracing out over the field degrees of freedom in the same way as it is routinely done for atoms in free space~\cite{lehmberg1970a,lehmberg1970b,agarwal1974,asenjo2017,peter2024} yields a dissipative coupling between the atoms described by an effective non-Hermitian interaction potential $U =  -(\hbar \Gamma_0/2) \sum_{a b} G_{a b}^{(\text{WG})} \sigma_{a 0}^+ \sigma_{b 0}^- $ describing exchange of excitations between different atoms mediated by the planar waveguide (WG) Green's matrix
\begin{equation}
    \begin{aligned}
    G_{ab}^{(\text{WG})} &= 
        -\frac{6\pi}{k_0} {\cal G}_{zz} (\vec{r}_{ab})
        = -\frac{6\pi}{k_0}
        \sum^{\infty}_{n = - \infty} {\cal G}^{(0)}_{zz} (\vec{r}_{ab} + n d \hat{\mathbf{z}}).
    \end{aligned}
    \label{eq:Green}
\end{equation}
Here $\vec{r}_{ab} = \vec{r}_b-\vec{r}_a$, $k_0 = \omega_0/c$ with $c$ the speed of light, $\Gamma_0$  is the spontaneous decay rate of the excited state in the free space and 
\begin{equation}
\begin{aligned}
{\cal G}^{(0)}_{ kl}(\vec{r}) 
&= -\frac{e^{i k_0 r}}{4\pi r}
\left[
    P (i k_0 r) \delta_{kl} 
    + Q (i k_0 r) \frac{r_{k} r_{l}}{r^2} 
\right]
+ \frac{\delta(\vec{r})}{3 k_0^2} \delta_{kl}
\end{aligned}
\label{eq:Green0}
\end{equation}
is the free-space dyadic Green's function of 3D Maxwell equations, where $P(u)=1-1/u+1/u^2$, $Q(u)=-1+3/u-3/u^2$, and $k, l = x, y ,z$.
In Eq.~(\ref{eq:Green}), the term
$-(6\pi/k_0) {\cal G}^{(0)}_{zz} (\vec{r}_{ab})
= G_{ab}^{(\text{3D})}$
corresponding to $n=0$ describes the direct propagation of light from $\vec{r}_a$ to $\vec{r}_b$ in the free 3D space, in the absence of the waveguide. We regularize its divergence for $a=b$ by redefining $\omega_0$: $\omega_0 \to \omega_0 + (\Gamma_0/2) (6\pi/k_0)\text{Re}{\cal G}^{(0)}_{zz}(0)$. Terms corresponding to $n \ne 0$ describe reflections of light at the mirrors defining the waveguide and ensure that the Green's function ${\cal G}_{zz} (\vec{r})$ obeys Neumann boundary conditions $\partial{\cal G}_{zz}(\vec{r})/\partial z = 0$ at $z = \pm d/2$.
Strictly speaking, the Green's function depends on the frequency $\omega = c k$ of the light field considered.Since atoms are resonant scatterers, their response to the field will be sizable only for $\omega \approx \omega_0$, that is for quasi-resonant light. In this case, one can replace $k$ by $k_0$ and obtain Eq.~\eqref{eq:Green0} provided $|kr-k_0r| \ll 1$. This will be the case for all inter-atomic distances as long as $|kR-k_0R| \ll 1$. This condition can be conveniently rephrased as the small detuning condition $|\omega-\omega_0|/\Gamma_0 \ll Q/(k_0 R)$ where $Q = \omega_0/\Gamma_0$ is the quality factor of the atomic resonance~\cite{sm}. For optical transitions, $Q$ is generally of the order of $10^7$--$10^8$.

\quad
In the single-excitation sector relevant for Anderson localization and in the quasi-resonant approximation, the total effective Hamiltonian $H_\mathrm{eff}=H_0+U$ provides an exact projection of the master equation for the atomic subsystem,
obtained after adiabatically eliminating the field~\cite{antezza_castin,antezza09}.
However, this projection discards so-called free-field solutions that are identically zero exactly at atomic positions. Previously discussed in Refs.\ \cite{klug06,antezza09} for periodic arrangements of atoms, such solutions can be excluded in a random system at least when the typical inter-atomic distance becomes much shorter than the resonant wavelength, leading to the condition $\rho/k_0^2 \gg 1/(2\pi)^2 \simeq 0.03$, which is obeyed by the atomic densities we consider in the following.

\quad Let us now study the right quasi-normal modes~\cite{note_eff_hamiltonian} $|\Psi_n\rangle$ of $H_\mathrm{eff}$ describing elementary excitations in the atomic ensemble defined above:  $H_\mathrm{eff} |\Psi_n \rangle = \hbar\left(\omega_n-i\Gamma_n/2\right)|\Psi_n\rangle$, and compare their properties with those of the modes in the pure 2D geometry, where~\cite{maximo_spatial_2015}
\begin{equation}
    \begin{aligned}
    G_{ab}^{(\text{2D})} &=\frac{2 \delta_{ab}}{\Gamma_0} \left[-\Delta \omega^{(\text{WG})} + i \frac{\Gamma^{(\text{WG})}}{2} \right]
\\    
&+ \frac{3i\pi  (1 - \delta_{ab})}{2k_0 d}H_0^{(1)}(k_0 r_{ab})
    \end{aligned}
\label{eq:Green2D}
\end{equation}
with $H_0^{(1)}$ the zeroth-order Hankel function of the first kind. $\Delta \omega^{(\text{WG})}$ and $\Gamma^{(\text{WG})}$ account for the modification of the atomic resonance frequency and of the decay rate of excited state, respectively, due to the presence of the waveguide, whereas the factor $1/k_0 d$ in front of $H_0$ compensates for the increase of the density of electromagnetic states with decreasing $d$~\cite{sm}.
Remarkably,
$\text{Im} G_{ab}^{(\text{WG})} = \text{Im} G_{ab}^{(\text{2D})}$ for all $r_{ab} > 0$, but $\text{Re} G_{ab}^{(\text{WG})} \simeq \text{Re} G_{ab}^{(\text{2D})}$ only for $k_0 r_{ab} > 1$, whereas $\text{Re} G_{ab}^{(\text{WG})} \simeq \text{Re} G_{ab}^{(\text{3D})}$ for $k_0 r_{ab} < 1$~\cite{sm}. These relations underpin both the similarities and the differences between modes of planar WG and 2D models---Eqs.~(\ref{eq:Green}) and (\ref{eq:Green2D}), respectively,---that we discuss below. A crucial point is that $\text{Re} G_{ab}^{(\text{3D})}$, and hence $\text{Re} G_{ab}^{(\text{WG})}$, both diverge like $1/r_{ab}^3$ for $r_{ab} \to 0$ whereas $\text{Re} G_{ab}^{(\text{2D})} \propto
-\ln(r_{ab})$ for $r_{ab} \to 0$. This makes coupling between neighboring atoms much stronger and can be mathematically understood as a cumulative contribution of evanescent modes of the empty waveguide that sum up to produce the $1/r_{ab}^3$ divergence of the Green's function~\cite{sm}. The resulting effective interaction $U$ is scalar only in the sense that a single polarization is involved. $U$ does not derive from a scalar wave equation. Instead, restricting the dynamics to
$s$ polarization selects one exact sector of the full
vectorial problem without discarding or artificially introducing
any near-field terms. The boundary conditions, however, shape the far-field Green's function in the waveguide to match the 2D scalar description. The latter description, as well as its 3D counterpart~\cite{skipetrov_absence_2014}, should be regarded as an idealized reference model, strictly justified only in certain regimes (\textit{e.g.}, at low atomic density). This is in contrast to Eq.~\eqref{eq:Green}, which derives from Maxwell's equations without any approximation.

\begin{figure}
    \centering
    \includegraphics[width=1.0\columnwidth]{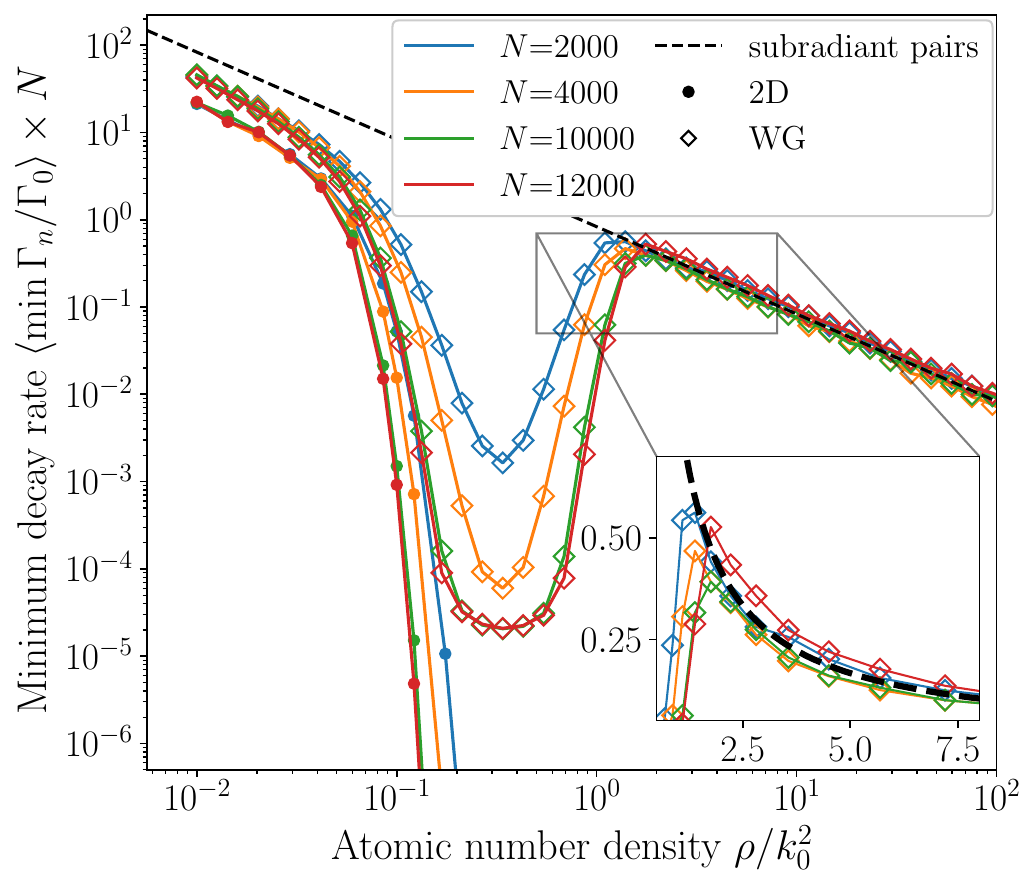}
    \caption{Ensemble average of the decay rate of the longest-lived quasi-normal mode, as a function of the atomic density for different system sizes, for modes of our planar WG system (open diamonds) and for truly 2D modes (dots). The black dashed line corresponds to the decay rate of the most subradiant mode localized on a pair of atoms~\cite{sm}.
    }\label{fig:min_im}
\end{figure}
 
\quad Let us now probe spatial localization of eigenvectors $|\Psi_n \rangle$ of $H_{\text{eff}}$ using three complementary diagnostics of
increasing sophistication: the quasi-mode decay rates $\Gamma_n$, inverse participation ratios and fractal dimensions of $|\Psi_n \rangle$, and, finally, a finite-size scaling analysis of the Thouless conductance. Figure~\ref{fig:min_im} shows the ensemble average of the minimum decay rate as a function of atomic number density $\rho$ for modes of both WG (open diamonds) and 2D (dots) models. For concreteness, we set $k_0 d = 0.9\pi$ from here on. 
For low densities $\rho/k_0^2 \lesssim 0.2$, all curves decrease rapidly with $\rho$ due to the phenomenon of Anderson localization resulting in $\langle \min \Gamma_n \rangle \propto \exp(-R/\xi)$~\cite{skipetrov_magnetic-field-driven_2015}, where $\xi$ is the localization length that decreases with $\rho$ faster than $R = {(N/\pi \rho)}^{1/2}$. This behavior is expected for scalar waves in 2D from the scaling theory of localization, predicting that all eigenmodes are localized at large $R$~\cite{Gang4}. Our results obtained for the pure 2D model consistently follow this prediction at all densities, with $\langle \min \Gamma_n \rangle$ becoming smaller than the numerical precision of our calculations for $\rho/k_0^2 \gtrsim 0.2$. In contrast, $\langle \min \Gamma_n \rangle$ obtained for atoms in a planar waveguide  exhibits a very different and unexpected behavior in this high-density regime. Instead of decreasing, it grows with $\rho$,
reaches a local maximum, and then scales as $\langle \min \Gamma_n \rangle/\Gamma_0 \propto k_0^2/(N \rho)$ for $\rho/k_0^2 \gtrsim 2$. The latter scaling corresponds to the average smallest decay rate associated with two-atom subradiance~\cite{sm}. Two-atom subradiant quasi-modes are localized on pairs of closely located atoms and are weakly sensitive to the presence of other atoms. The fact that such quasi-modes have the smallest decay rates among all quasi-modes for $\rho/k_0^2 \gtrsim 2$, signals the disappearance of long-lived Anderson-localized modes at such high densities of atoms.    

\begin{figure}
    \centering
    \includegraphics[width=1\columnwidth]{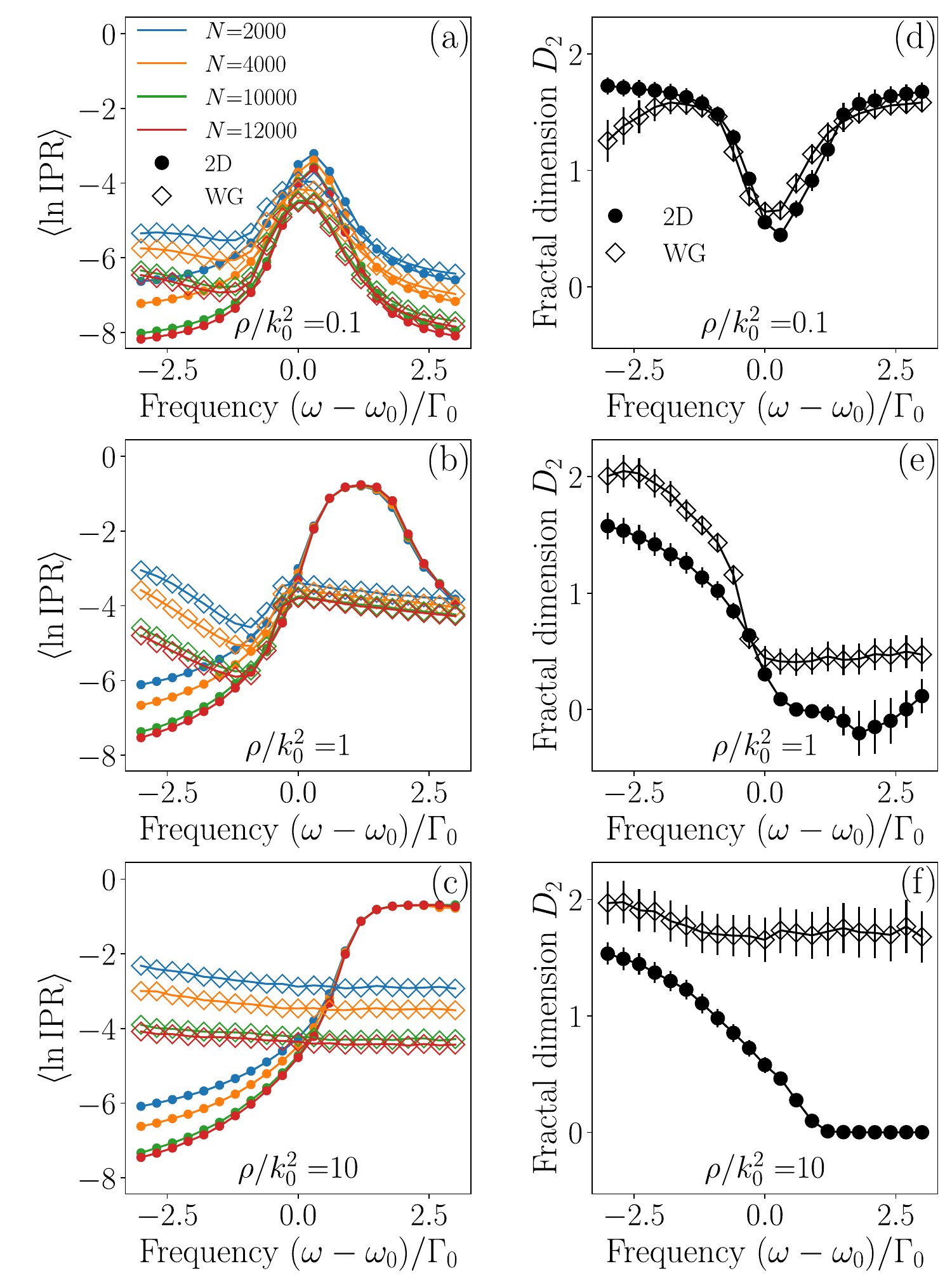}
    \caption{Mean-log inverse participation ratio $\langle \ln \mathrm{IPR} \rangle$ (a\textendash c) and fractal dimension $D_2$ (d\textendash f) as functions of frequency for different densities $\rho/k_0^2=0.1$ (top), $\rho/k_0^2=1$ (middle), $\rho/k_0^2=10$ (bottom) and different atom numbers $N$, for modes of both WG (open diamonds) and 2D (dots) models.}
    \label{fig:results_eigen_vect}
\end{figure}

\quad
To assess the localization of quasi-modes $|\Psi_n \rangle$ in a more direct way, we evaluate their inverse participation ratios $\text{IPR}_{n} = \sum_{a} \left| \Psi_n(\vec{r}_a) \right|^{4}$ and multifractal dimension $D_2$ defined by $\langle \ln \text{IPR} \rangle \propto - D_2
\, \ln R$~\cite{evers08,sm}.
From here on, $\langle \cdots \rangle$ denotes averaging over an ensemble of random atomic configurations $\{ \vec{r}_a \}$ and a moving average over a frequency band of width $\Gamma_0/4$, with the latter helping to reduce residual statistical fluctuations.
For modes localized on $M < N$ atoms, we expect $\text{IPR}_{n} \sim 1/M$ and $D_2 = 0$. In contrast, extended modes should have $\text{IPR}_{n} \sim 1/N$ and $D_2 = 2$. Figure~\ref{fig:results_eigen_vect}
displays $\langle \ln  \mathrm{IPR} \rangle$ and $D_2$ as functions of frequency for modes of both 2D and WG models for different densities and system sizes, in a frequency range where Anderson localization effects are the strongest.
For both 2D and WG cases, $\langle \ln  \mathrm{IPR} \rangle$ and $D_2$ behave similarly and exhibit signs of localization (increased $\langle \ln  \mathrm{IPR} \rangle$ and $D_2 $ significantly less than 2) at a low density [Fig.~\ref{fig:results_eigen_vect}(a) and (d)]. However, WG and 2D results start to differ at an intermediate density [Fig.~\ref{fig:results_eigen_vect}(b) and (e)] and become qualitatively different at high density [Fig.~\ref{fig:results_eigen_vect}(c) and (f)]. In the latter case, WG modes are characterized by low $\langle \ln  \mathrm{IPR} \rangle$ that decreases with $N$ and $D_2$ close to 2, which classifies them as extended. This is in contrast to 2D modes, which consistently exhibit high $\langle \ln  \mathrm{IPR} \rangle$ and $ D_2 \to 0$ at frequencies around $(\omega-\omega_0)/\Gamma_0 \sim 1$, for which signatures of Anderson localization are the strongest. Higher-order participation ratios and fractal dimensions behave in a qualitatively similar way~\cite{sm}.

\quad   

To prove that low- and high-density behaviors discussed above are separated by a sharp transition at some critical density $\rho_c$, we perform a finite-size scaling analysis.
We define the Thouless conductance $g_n = \delta\omega_n/\Delta\omega_n$ with the mode spectral width $\delta\omega_n = \Gamma_n$ and the spacing between ordered mode frequencies $\Delta\omega_n = |\omega_{n+1} - \omega_{n}|$~\cite{thouless1977,Gang4,wang11,skipetrov_absence_2014}.
Frequency ranges in which quasi-modes are localized in the limit of infinite sample size can be identified by analyzing the dependence of $\langle \ln g \rangle$ on sample size $R$. The scaling function $\beta= d \langle \ln g \rangle/d\ln k_0 R$ is predicted negative (resp. positive) in the frequency ranges where modes are localized (resp. extended)~\cite{Gang4}.

\quad As expected from the single-parameter scaling hypothesis, the scaling function calculated for the 2D model (\ref{eq:Green2D}) depends on a single parameter $\langle \ln g \rangle$ and remains negative within the accuracy of our calculations, in agreement with the localized nature of all modes~\cite{sm}. However, for quasi-modes following from Eq.~(\ref{eq:Green}) this is only true at low densities $\rho/k_0^2 \lesssim 0.5$~\cite{sm}. At higher densities, the single-parameter scaling breaks down and $\beta$ becomes function of both $\langle \ln g \rangle$ and $\rho$. Nevertheless, the sign of $\beta$ can still be used to conclude about spatial localization of quasi-modes in the large-sample limit. Figure~\ref{fig:scaling_tm}(a) clearly shows that whereas $\beta < 0$ at least at some frequencies for low densities $\rho/k_0^2 < 2$, it becomes positive at all frequencies when $\rho/k_0^2 \gtrsim 2$, signaling a localization-delocalization transition upon increasing $\rho$. 
At large densities, Fig.~\ref{fig:scaling_tm}(a) exhibits two frequency ranges with nearly constant $\beta$, one with $\beta\simeq 1$ and the other with $\beta\simeq 2$. The former, marked with left-leaning hatching, lies in the positive-detuning region ($\omega-\omega_0>0$) and is indicative of ballistic transport. The latter, marked with right-leaning hatching, lies in the negative-detuning region ($\omega-\omega_0<0$) and corresponds to subradiant modes localized on pairs of closely spaced atoms. Our scaling analysis clearly distinguishes these subradiant modes from the genuine Anderson-localized modes that yield $\beta <0$ (see also Ref.~\cite{sm}).       

\begin{figure}[!t]
    \centering
    \includegraphics[width=1.0\columnwidth]{Figures/critical_scaling.png}
    \caption{(a) Scaling function $\beta= d \langle \ln g \rangle/d\ln k_0 R$ for WG model (\ref{eq:Green}) versus frequency $\omega$ and atomic number density \(\rho\). $\beta$ is numerically computed from the dependence of $\langle\ln g  \rangle$ on the atomic sample size $R$ by a finite-difference approximation for $N = 4000$ and $10000$. The dashed black line indicates $\beta=0$. The large values $\beta > 2$ observed in some regions of the parameter plane are due to finite-size effects~\cite{sm}. The hatched high-density domain at $\omega-\omega_0 > 0$ corresponds to
    $\beta\simeq 1$ while the hatched high-density domain at $\omega-\omega_0 < 0$ corresponds to $\beta\simeq 2$.
    (b) $\langle \ln g \rangle$ obtained from the data for $(\omega-\omega_0)/\Gamma_0\in[0,3]$,  $N = 4000, 8000, 10000, 12000$ and $\rho/k_0^2\in [1,3]$ as a function of $R/\xi$  (symbols), and the best fit to it (solid line). The inset shows $\langle \ln g \rangle$ as a function of the scaling variable $u$. 
    (c) Frequency-resolved estimation of the critical exponent $\nu$ (symbols) compared to its value obtained from the global fit for all frequencies (solid black line) with uncertainty shown by gray dashed lines (see also Ref.~\cite{sm}).
    }
    \label{fig:scaling_tm}
\end{figure}

\quad We now focus on a portion of the critical line $\beta = 0$ indicated in green in Fig.~\ref{fig:scaling_tm}(a): $\rho/k_0^2\approx2$ for $\omega_0 < \omega < \omega_0 + 3 \Gamma_0$. We assume that in the vicinity of this line, $\langle \ln g \rangle$ becomes function of a single parameter $R/\xi$: $\langle \ln g \rangle = f(R/\xi)$ (single-parameter scaling hypothesis) and that the localization length diverges as $\xi = \xi_0 |\rho/k_0^2-\rho_c/k_0^2|^{-\nu}$, where $\nu$ is the critical exponent~\cite{Slevin2014}. Introducing a scaling variable $u=(k_0R)^{1/\nu}(\rho-\rho_c)/k_0^2$ and allowing for frequency dependence of the critical density $\rho_c(\omega)=\rho_c(\omega_0) + [(\omega-\omega_0)/\Gamma_0]\delta\rho_c$, we expand $\langle \ln g \rangle$ to second order in $u$: 
$\langle \ln g \rangle = c_0 + c_1 u + c_2 u^2$, and determine $\rho_c$, $\delta\rho_c$, $\nu$, $c_0$, $c_1$, and $c_2$ from the best fit to numerical data corresponding to different $R$, $\rho$ and $\omega$ using the trust region reflective algorithm~\cite{trust_ref_algo}. We find in particular $\nu = 1.4 \pm 0.2$ (see Ref.~\cite{sm} for a detailed discussion of the fitting procedure, statistical uncertainties, and corrections to scaling).
Finally, we validate the single-parameter scaling hypothesis, by collapsing all our numerical data on a master curve $\langle \ln g \rangle = f(R/\xi)$ obtained as a result of the fit, see Fig.~\ref{fig:scaling_tm}(b). The collapse remains stable upon moderate variation of the frequency band used for the fit.

\quad To test the universality of the critical exponent, we repeat the fits for each frequency independently, allowing for arbitrary dependence $\rho_c(\omega)$ and different $\nu$ for different $\omega$. The best-fit $\nu$ shown by circles in Fig.~\ref{fig:scaling_tm}(c) turn out to be frequency independent within the accuracy of our analysis, confirming the universality of $\nu$. They also agree well with the estimate obtained from the global fit performed for all frequencies at once (black horizontal line).  

\quad In conclusion, we have discovered a localization-delocalization transition for $s$-polarized light scattered by a 2D disordered ensemble of resonant point scatterers (resonant wave number $k_0$) in a thin planar waveguide of width $d < \pi/k_0$. The transition is driven by the strengthening of quasi-static dipole-dipole interactions between scatterers upon increasing their areal number density $\rho$---a mechanism similar to the one preventing localization in 3D~\cite{skipetrov_absence_2014}. However, in contrast to 3D where no localized modes were found at any density, the considered 2D system features a genuine transition from Anderson localization at low $\rho$ to a different transport regime at high $\rho$, with a mobility edge at a critical density $\rho_c \simeq 2 k_0^2$ (for $k_0 d = 0.9\pi$) and a critical exponent  $\nu=1.4 \pm 0.2$. Whereas $\rho_c$ may depend on $d$, $\nu$ is expected to be universal. It remains to be seen whether this transition falls under the standard ten-fold Altland-Zirnbauer classification~\cite{zirnbauer96,altland97,evers08} and what is its universality class. Transport properties of high-density phase replacing Anderson localization for $\rho > \rho_c$ will also be a subject of future studies. Finally, a similar transition may be expected for resonant point scatterers in a 1D waveguide as well as for other types of resonant scatterers (e.g., small dielectric spheres) in both 1D and 2D, provided that high scatterer densities can be reached.

\nocite{milonni_spontaneous,retarded, Sokolov2011, wulles_honey,bettles_cooperative_2016,skipetrov_eigenvalue_2011,efron_1979,bevington2002,slevin_ohtsuki_crit_AL,MullerDelande2010,brown2011fourier,lagendijk_resonant_1996,modal_complexity}

\putbib
\end{bibunit}

\clearpage

\onecolumngrid

\begin{center}
    \Large{\bf{Supplementary material for ``Delocalization transition for light in two dimensions''}}
\end{center}

\onecolumngrid

\begin{center}
\parbox{0.75\textwidth}{\small Here we present (i) a detailed analysis of the electromagnetic Green’s function in a planar waveguide and its relation to 2D and 3D free-space cases, as well as single-atom radiative properties; (ii) an analytical estimate of subradiant decay rates based on nearest-neighbor statistics; (iii) the statistical analysis of the Thouless conductance and its uncertainty; (iv) the procedure used to extract the critical exponent, including fitting methods and error estimation; (v) a comparison with Anderson localization at low density and discussion of finite-size effects; (vi) an analytical derivation of conductance in the diffusion regime; (vii) an analysis of inverse participation ratios and fractal dimensions at order $q\ge2$; and (viii) representative mode profiles illustrating localized and extended states, and numerical evidence that localization of collective atomic excitations implies localization of the electromagnetic field.}
\end{center}

\twocolumngrid

\section{Green's function in a planar waveguide}

\quad Throughout this work we focus on the regime in which the distance $d$ between the two mirrors of a planar waveguide satisfies $k_0d < \pi$, where $k_0$ is the optical wavenumber associated with the single-atom resonance frequency $\omega_0$ in the free space. In this case, only transverse-magnetic modes TM$_{0}$ (electric field perpendicular to the mirrors) can propagate in the waveguide (i.e., have real wavevectors). For concreteness, we put $k_0d=0.9\pi$ in our numerical calculations. In this section, we compare the Green’s function describing the coupling between two atoms placed in the middle plane $z=0$ of the waveguide (WG) and polarized perpendicular to the atomic plane (the so-called $s$ polarization~\cite{maximo_spatial_2015}) with the free-space Green’s functions of 2D scalar and 3D vectorial problems. For $r>0$, the three Green's functions, that are properly normalized to allow for their direct comparison, are
\begin{eqnarray}
        \tilde{\cal G}^{(\text{2D})}(\vec{r}) &=& \frac{3i\pi}{2k_0 d} H_0^{(1)}(k_0 r)
        \label{green2d}
         \\
         \tilde{\cal G}^{(\text{3D})}(\vec{r}) &=& -\frac{6\pi}{k_0} {\cal G}_{zz}^{(0)} (\vec{r}) 
        \label{green3d}
        \\
         \tilde{\cal G}^{(\text{WG})}(\vec{r}) &=& -\frac{6\pi}{k_0} \sum^{\infty}_{n = - \infty} {\cal G}^{(0)}_{zz} (\vec{r} + n d \hat{\mathbf{z}})
        \label{greenwg}
\end{eqnarray}
Here $H_0^{(1)}$ the zeroth-order Hankel function of the first kind and
\begin{equation}
    {\cal G}^{(0)}_{ \mu\nu}(\vec{r}) 
    = -\frac{e^{i k_0 r}}{4\pi r}
    \left[
        P (i k_0 r) \delta_{\mu\nu} 
        + Q (i k_0 r) \frac{r_{\mu} r_{\nu}}{r^2} 
    \right]
    + \frac{\delta(\vec{r})}{3 k_0^2} \delta_{\mu\nu}
\end{equation}
is the free-space dyadic Green's function of 3D Maxwell equations, $\mu, \nu = x, y ,z$, $P(u)=1-1/u+1/u^2$, and $Q(u)=-1+3/u-3/u^2$. A multiplicative factor $(k_0 d)^{-1}$ in Eq.~(\ref{green2d}) compensates for the increase of the density of electromagnetic states as $d$ decreases. Strictly speaking, the Green's function depends on the frequency $\omega = ck$ of the light considered, leading to a non-linear self-consistent set of coupled equations. Throughout this work, we adopt the standard quasi-resonant approximation and evaluate it at the atomic transition frequency $k\approx k_0$. This so-called \textit{polar approximation} is valid provided the frequency dependence of the Green's function is negligible over the relevant spectral range~\cite{Sokolov2011,retarded}. Since both in the waveguide and in the pure 2D model, the Green's function in the far-field ${\mathcal{G}} (r, \omega)\propto e^{ikr} /
\sqrt{kr}$, expanding around the atomic resonance yields
\begin{align}
  \gamma = \left| \frac{\mathcal{G}(r, \omega) - \mathcal{G}(r, \omega_0)}{\mathcal{G}(r, \omega_0)} \right| &
  \approx 
 |\delta \omega| \frac{r}{c}
 <  |\delta \omega| \frac{2R}{c}
\end{align}
where $\delta \omega  = \omega-\omega_0$. To check whether the condition $\gamma \ll 1$ is satisfied in our calculations, we rewrite it in terms of the variables we use as $|\delta \omega|/\Gamma_0 \ll Q/k_0R$,
where $Q = \omega_0 / \Gamma_0$ is the quality factor of the atomic resonance (typically, $Q \sim 10^7$--$10^8$). The condition $\gamma \ll 1$ is then satisfied with a wide margin for all calculations that we present in the manuscript. 

\begin{figure}
    \centering
    \includegraphics[width=1.0\columnwidth]{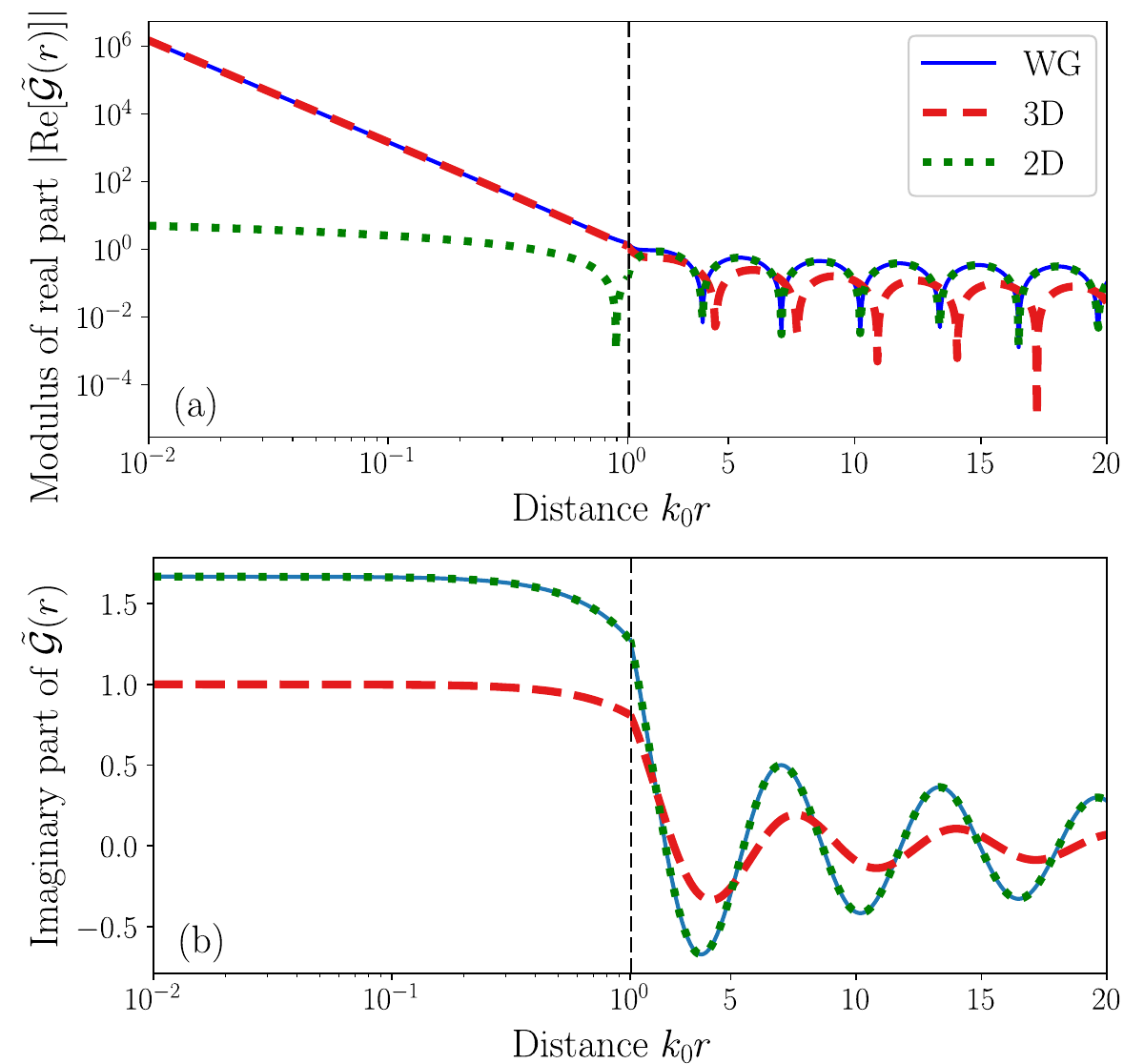}
    \caption{Real (a) and imaginary (b) parts of the Green's function $ \tilde{\cal G}^{(\text{WG})}$ inside the waveguide for $k_0d/\pi=0.9$, as functions of the distance $r$ (blue solid line). $ \tilde{\cal G}^{(\text{WG})}$ is compared to the 2D scalar Green's function $ \tilde{\cal G}^{(\text{2D})}$ (green dotted line) and the 3D Green's function $ \tilde{\cal G}^{(\text{3D})}$ (red dashed line). Note the change from logarithmic to linear scale of the horizontal axis at $k_0r=1$, indicated by the black vertical dashed line.
}\label{fig:green_TM}
\end{figure}

\quad Figure \ref{fig:green_TM} shows that the Green's function in the waveguide coincides with the 2D free-space one for long distances $r > 1/k_0$, whereas for short distances $r < 1/k_0$ only its imaginary part behaves in the same way as in 2D. The real part of $ \tilde{\cal G}^{\text{(WG)}}$ becomes equal to its 3D counterpart instead, diverging as $1/r^3$ for $r \to 0$, see Fig.~\ref{fig:green_TM}(a). This is the main microscopic factor underlying the results discussed in the main text.

\begin{figure}
    \centering
    \includegraphics[width=1.0\columnwidth]{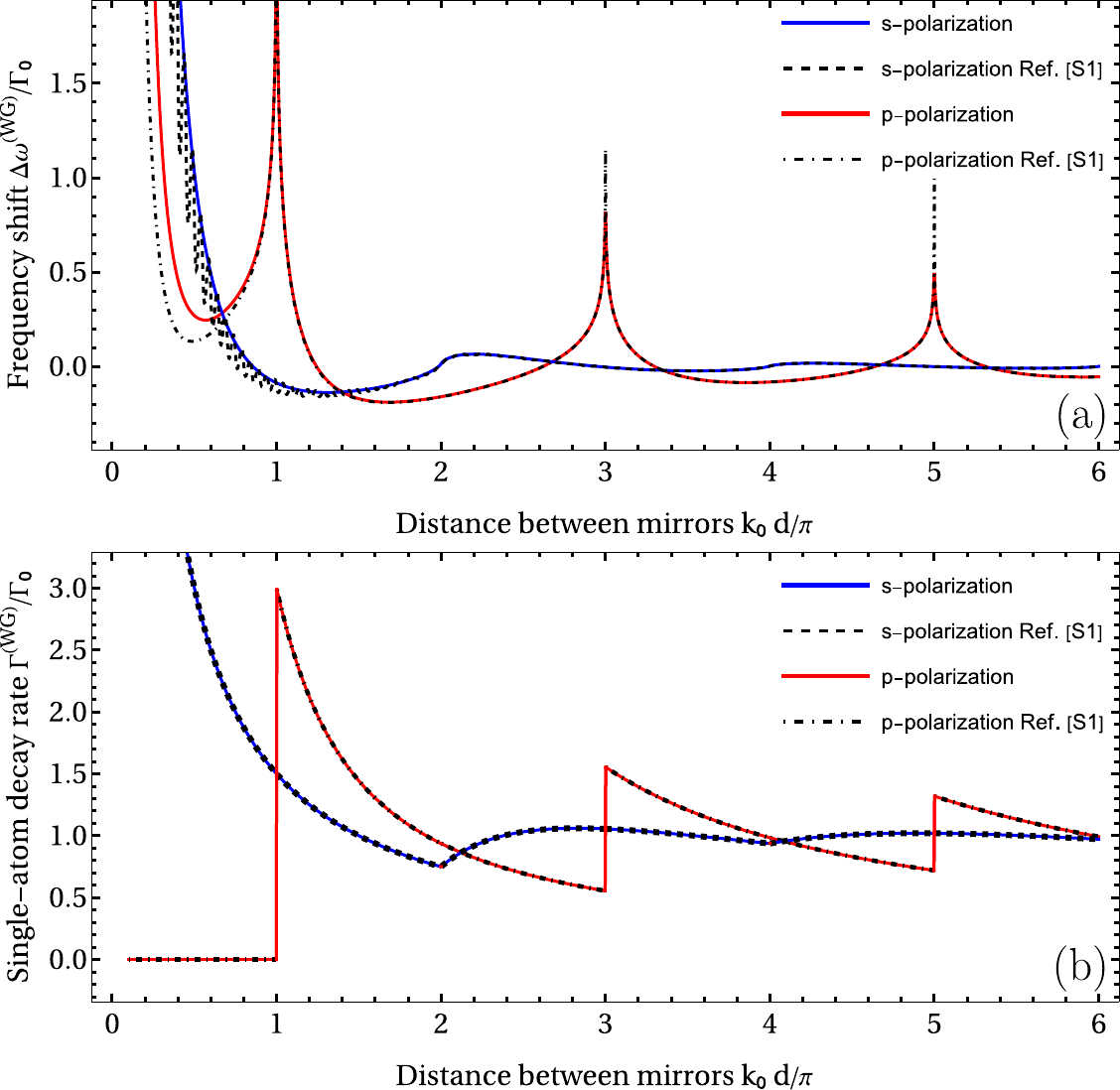}
    \caption{Properties of a single, isolated atom located in the middle plane of a planar 2D waveguide:  the resonance frequency shift $\Delta\omega^{\text{(WG)}}$ (a) and the decay rate of the excited state $\Gamma^{\text{(WG)}}$ (b), as functions of the distance $d$ between the waveguide mirrors for both an $s$-polarized dipole interacting with TM modes (blue) and a $p$-polarized dipole interacting with TE modes (red). We compare our calculations with those of Ref.~\cite{milonni_spontaneous} (black dashed and dot-dashed lines), where the longitudinal part of the dyadic Green's function has been neglected.
}\label{fig:green_tm_0}
\end{figure}

\quad The waveguide also affects radiation properties of a single, isolated atom, namely its resonance frequency and decay rate. The shift of the resonance frequency $\Delta \omega^{(\text{WG})}$ and the decay rate $\Gamma^{(\text{WG})}$ are obtained from the limit of $ \tilde{\cal G}^{(\text{WG})}(\vec{r})$ for $r \to 0$ as
\begin{equation}
    \begin{aligned}
        \Delta \omega^{(\text{WG})} &= -\frac{\Gamma_0}{2}  \mathrm{Re}\left\{
        \lim_{r \to 0}
        \left[
         \tilde{\cal G}^{(\text{WG})}(\vec r) -  \tilde{\cal G}^{(\text{3D})}(\vec r)
        \right]
        \right\}\\
        \Gamma^{(\text{WG})} &= \Gamma_0 \mathrm{Im}\left[
        \lim_{r \to 0}
         \tilde{\cal G}^{(\text{WG})}(\vec r)
        \right]
    \end{aligned}
    \label{limits}
\end{equation}
where $\Gamma_0$  is the spontaneous decay rate of the excited state in the free space. Similar results can be also obtained for the in-plane (so-called $p$) atomic polarization~\cite{milonni_spontaneous}.

\quad The limits in Eqs.~(\ref{limits}) are calculated in the same way as in Refs.~\cite{wulles_honey,bettles_cooperative_2016}. The resulting $\Delta \omega^{(\text{WG})}$ and $\Gamma^{(\text{WG})}$ are shown in Fig.~\ref{fig:green_tm_0} for atoms with both out-of-plane ($s$-polarized) and in-plane ($p$-polarized) dipole orientations, interacting respectively with the transverse magnetic (TM) and transverse electric (TE) modes of the waveguide. A comparison with previous calculations \cite{milonni_spontaneous} highlights the importance of properly accounting for the longitudinal part of the Green's function when computing the shift of the resonance frequency for small $k_0 d < \pi$, see Fig.~\ref{fig:green_tm_0}(a). The decay rate turns out to be insensitive to the longitudinal part of the Green's function, see Fig.~\ref{fig:green_tm_0}(b).  

\begin{figure}
\centering{   
\includegraphics[width=\columnwidth]{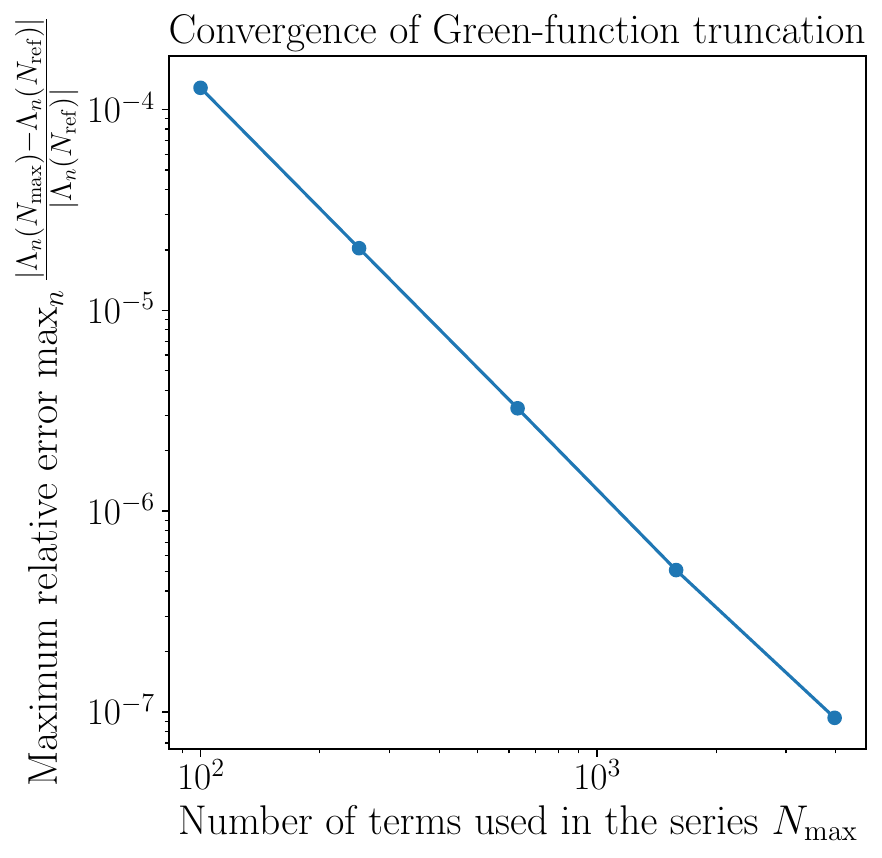}
}
    \caption{
      Maximum relative error $\max_n \{ |\Lambda_n(N_{\text{max}})-\Lambda_n(N_\mathrm{ref})|/|\Lambda_n(N_\mathrm{ref})|\}$ of the spectrum $\{\Lambda_n\}$ of the effective Hamiltonian, as a function of the number of terms $N_\mathrm{max}$ retained in the series Eq.~\eqref{greenwg} compared to the computation with $N_\mathrm{max}=N_\mathrm{ref}=10^4$. The calculation is performed for $N=150$ atoms at a density $\rho / k_0^2 = 1$.
    }\label{fig:green_trunc}
\end{figure}

\quad In order to evaluate numerically the Green's function in the waveguide, we truncate the series in Eq.~\eqref{greenwg} symmetrically at $| n | = N_{\max}$. Although the asymptotic behavior of $\mathcal{G}^{(0)}_{\mathrm{zz}} (\mathbf{r} + n d \mathbf{z}) =\mathcal{O} (e^{i k_0 d | n |} / | n | d)$ is not absolutely converging, the oscillatory phase factor ensures rapid convergence of the symmetrically truncated series for the parameters considered. In practice, we set $N_{\max} = 10^4$ for which all eigenvalues of the effective Hamiltonian $H_{\mathrm{eff}}$ have converged to their true values with an accuracy of at least $10^{- 8}$ (relative tolerance) in all regimes considered. Figure~\ref{fig:green_trunc} shows an example of convergence as the number of terms in the sum of Eq.~\eqref{greenwg} increases. To accelerate the numerical evaluation of the series, the Green's function was computed using a hybrid scheme combining the exact series representation for intermediate distances with analytic short-range and long-range asymptotics. The intermediate regime was precomputed on a dense grid in separation distance and stored. During the construction of the matrix representation of $H_{\mathrm{eff}}$, values were obtained by interpolation. We verified that the resulting interpolation and truncation errors are negligible by systematically comparing to direct summation with increasing $N_{\max}$, as well as to the asymptotic forms in their respective validity regimes.

\section{Two-atom subradiance}\label{app:subrad_pairs}

\quad In this section, we derive an analytical expression for the average value of the minimal decay rate $\langle \min \Gamma_n \rangle$ of WG quasi-modes at high atomic number density $\rho$, adapting the approach  of Ref.~\cite{skipetrov_eigenvalue_2011}. Numerical simulations show that in the absence of Anderson localization, the minimum decay rate is reached by a subradiant mode localized on a pair of closest atoms in the sample. Two atoms separated by a distance $r$ give rise to super- and subradiant modes with decay rate $\Gamma_{\pm}(r) = \Gamma^{(\text{WG})} \pm \Gamma_0 \mathrm{Im}[ \tilde{\cal G}^{(\mathrm{WG})}(r)]$, respectively. For the subradiant mode and $k_0r \ll 1$, we have
\begin{equation}
    \begin{split}
        \Gamma_{-}(r)/\Gamma_0 &= \Gamma^{(\mathrm{WG})}/\Gamma_0 - \mathrm{Im}[ \tilde{\cal G}^{(\mathrm{WG})}(\vec{r})]\\
        &= \Gamma^{(\mathrm{WG})}/\Gamma_0 - \mathrm{Im}[ \tilde{\cal G}^{(\mathrm{2D})}(\vec{r})]\\
        &= \frac{3 \pi}{2 k_0d} (1 - \mathrm{Im}[i H^{(1)}_0(k_0 r)]) \\
    \end{split}
    \label{gammasub}
\end{equation}

\quad To find the minimum of Eq.~(\ref{gammasub}) averaged over all possible atomic configurations, we first compute the probability density that, given $N$ atoms uniformly distributed in a disk of area $S$ with density $\rho=N/S$, the two closest atoms are separated by a distance  $r$. Let us choose an arbitrary atom $a$. The probability that another atom $b$ is located in an annular shell of radius $r$ and thickness $dr$ around the first is $p_1 = 2\pi r dr/S$. For $r$ to be the minimal distance, all other $N - 2$ atoms have to be outside the disk of radius $r$ centered on $a$. The probability of this is $p_2 = {(1 - \pi r^2/S )}^{N-2}$. In addition, the distances between ${(N - 1)}^2$ pairs of atoms excluding the atom $a$ have to be larger than $r$, which has a probability $p_3 = {(1 - \pi r^2/S )}^{{(N-1)}^2}$. Thus, the probability that $r$ is the minimum distance between any two atoms is equal to the number of possibilities $N (N - 1)/2$ to choose the two atoms $a$ and $b$, multiplied by the product $p_1 p_2 p_3$. Then, the probability density is
\begin{equation}
    \begin{split}
        p(r) & = \frac{2\pi}{S} r N(N-1){\left(1-\frac{\pi r^2}{S}\right)}^{N(N-1)/2 - 1}\\
        &\simeq
        \pi\rho r N e ^ {-\pi r^2 N \rho / 2}
    \end{split}
\end{equation}
where the second line is obtained in the limit $N  \to \infty$. Finally, we have

\begin{equation}
    \begin{split}
         \langle \min \Gamma_n / \Gamma_0 \rangle & = \int_{0}^{\infty} d r p(r) \Gamma_{-}(r)\\
         & = \frac{3\pi}{2 k_0 d } \left( 1 - \exp{\left(- \frac{k_0^2}{2 \pi N \rho } \right)} \right) \\
         & \approx \frac{3k_0^2}{4 k_0d N \rho} 
    \end{split}
    \label{mingamma}
\end{equation}
\noindent where we expanded the final result to first order in $k_0^2/(N \rho)$.
The derivation assumes that, for each disorder realization, the minimum decay rate is given by the two-atom subradiant decay rate of the closest pair, \textit{i.e.}, $\min\Gamma_n=\Gamma_-$. While this relation holds for the majority of random configurations, there exist rare but statistically important configurations for which $\min\Gamma_n\neq\Gamma_-$. These configurations affect the ensemble average and therefore make the prefactor obtained in Eq.~\eqref{mingamma} inaccurate. Nevertheless, the treatment correctly captures the asymptotic scaling $\langle \min\Gamma_n\rangle\propto 3k_0/(4 d N_{\mathrm{at}} \rho)$. In the main text, we therefore compare the numerical results with this asymptotic scaling, allowing for an adjusted constant prefactor.

\section{Average logarithm of conductance and its uncertainty}\label{app:error}

\quad To analyze the conductance $g =\delta \omega/\Delta \omega$ as a function of frequency $\omega$, we define a set of $N_f$ regularly spaced frequencies $\omega_k$. For each $\omega_k$, $k=1, \hdots, N_f$, and for each system size $R_l$, $l=1, \hdots, N_R$ where $N_R$ is the number of system sizes considered, we compute the average logarithm of conductance $\langle \ln g \rangle(\omega_k,R_l)$ from $N_\mathrm{rea}$ independent realizations of the system. For each realization, we introduce $\mathcal{J}_r(\omega_k)
= \left\{ j \in \{1,\dots,N\} : |\omega_j - \omega_k| < \tfrac{\Delta}{2} \right\}$, of cardinality $n_r = |\mathcal{J}_r(\omega_k)|$, 
where the modes are sorted by increasing frequencies $\omega_j$ and $\Delta = \Gamma_0/4$. We then estimate the average logarithm of conductance as
\begin{equation}
    \begin{aligned}
        \langle \ln g \rangle(\omega_k,R_l) & = \frac{1}{N_\mathrm{rea}}\sum_{r=1}^{N_\mathrm{rea}} \ln g_r \\
        \ln g_r &= \ln\delta\omega_r - \ln\Delta\omega_r\\
        \ln \delta\omega_r &= \frac{1}{n_r}\sum_{j=1}^{n_r} \ln\Gamma_j \\
        \ln\Delta\omega_r &= \frac{1}{n_r-1}\sum_{j=1}^{n_r-1} \ln\left( \omega_{j+1} - \omega_{j} \right) \\
    \end{aligned}
\end{equation}

The variance of the average logarithm of conductance is given by the standard unbiased estimator: 
\begin{equation}
    \sigma^2(\omega_k, R_l) = {(N_\mathrm{rea}-1)}^{-1} \sum_{r=1}^{N_\mathrm{rea}} {\left[ \ln g_r - \braket{\ln g}(\omega_k, R_l) \right]}^2
\end{equation}
The uncertainty of estimation of $ \langle \ln g \rangle(\omega_k,R_l)$ is computed as $\sigma_{\mathrm{tot}}(\omega_k, R_l) = \sigma(\omega_k, R_l) / \sqrt{N_\mathrm{rea}}$.

\section{Critical exponent estimation}
\label{app:fitting}

\begin{figure*}
    \centering
    \includegraphics[width=\textwidth]{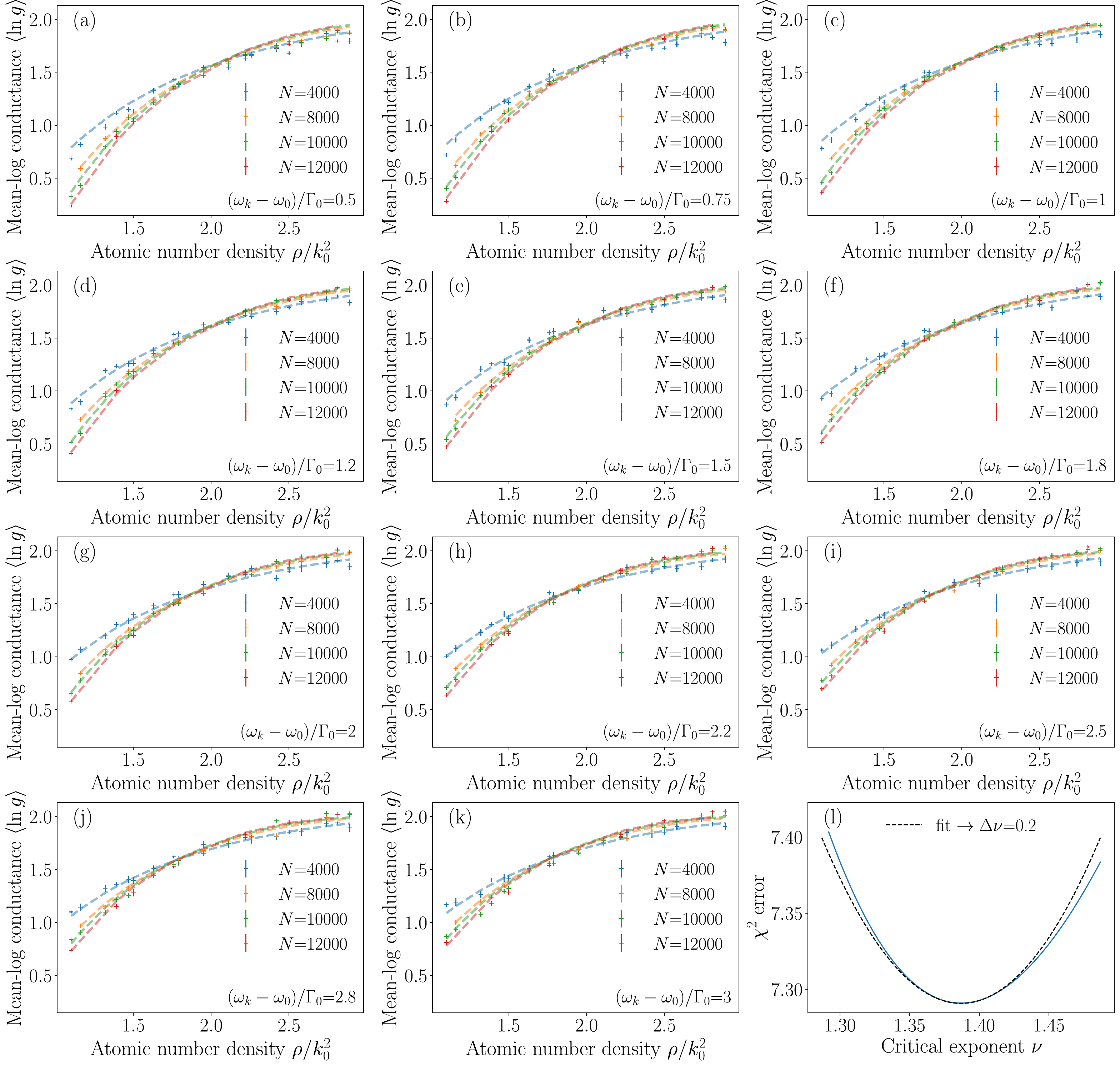}
    \caption{ (a\textendash k) Average logarithm of the conductance $\braket{\ln g}$ as a function of density, for different system sizes ('+' symbols with error bars computed as detailed in Sec.~\ref{app:error}), together with the corresponding fits (dashed lines).
    (l) $\chi^2$ statistics as a function of $\nu$, obtained by fixing $\nu$ and adjusting all other parameters to obtain the best fit (solid line). Dashed line shows the parabolic approximation (\ref{parabola}).
}
\label{fig:fit_quality}
\end{figure*}

\quad In this section we detail the estimation of the critical exponent $\nu$ described in the main text and the evaluation of its uncertainty. We choose frequency and density intervals $(\omega-\omega)/\Gamma_0 \in [0,3]$, $\rho/k_0^2 \in [1,3]$ to ensure having enough statistics to explore the vicinity of the density-driven localization-delocalization transition taking place around $\rho/k_0^2 \simeq 2$. We fit our numerical results for $\langle \ln g \rangle(\omega_k,R_l)$ in this frequency and density ranges by a function $F[u(\omega_k,R_l)] = c_0 + c_1 u + c_2 u ^2$ with a scaling variable $u(\omega_k,R_l)={(k_0R_l)}^{1/\nu}[\rho-\rho_c(\omega_k)]/k_0^2$, where the critical density $\rho_c(\omega) = \rho_c(\omega_0) + (\omega-\omega_0)\delta\rho_c$ is allowed to depend on the frequency $\omega$. The six free fit parameters are obtained by minimizing the $\chi^2$ statistic defined as
\begin{equation}\label{eq:residuals}
\chi^2 = \frac{1}{N_f N_R}\sum_{k,l} {\left\{ \frac{\braket{\ln g}(\omega_k, R_l) - F[u(\omega_k,R_l)]}{\sigma_{\mathrm{tot}}(\omega_k, R_l)}\right\}}^2.
\end{equation} 
The fit is performed using the trust region reflective algorithm implemented in the SciPy library~\cite{trust_ref_algo}. The best fits are shown in Figs.~\ref{fig:fit_quality}(a\textendash k), each panel corresponding to a different frequency $\omega_k$. The best-fit parameters are $\nu=1.4\pm0.1$, $\rho_c(\omega_0)/k_0^2= 2.1 \pm 0.3$, $\delta\rho_c/k_0^2=-0.122\pm0.003$, $c_0=1.56\pm0.02$, $c_1=0.051 \pm 0.006$ and $c_2=-0.0015\pm0.0004$ where uncertainties are estimated using standard bootstrap resampling methods~\cite{efron_1979}. 
The obtained estimation of $\nu$ is rather robust with respect to extending the frequency range used for the fit to higher frequencies. 

\quad To improve the estimation of the uncertainty of $\nu$, we analyze the $\chi^2$ landscape. We consider a range of candidate values denoted by $\{ \nu_i \}$ in the vicinity of the best-fit value $\nu_\mathrm{best} = 1.4\pm0.1$ obtained above. For each fixed $\nu_i$, all other parameters are adjusted to minimize $\chi^2$. By construction, the minimum of $\chi^2(\nu_i)$ identifies the best-fit exponent: $ \nu_\mathrm{best} = \arg \min_{\nu_i} \chi^2(\nu_i)$. Near its minimum, the $\chi^2$ curve is approximated by a parabola
\begin{equation}
\chi^2(\nu) \approx \chi^2_\mathrm{\min} + \left. \frac{1}{2} \frac{d^2 \chi^2}{d \nu^2}\right|_{\nu_\mathrm{best}} {(\nu - \nu_\mathrm{best})}^2.
\label{parabola}
\end{equation}
Figure~\ref{fig:fit_quality}(l) shows both $\chi^2(\nu)$ and its approximation (\ref{parabola}).
The standard error of $\nu_\mathrm{best}$ is then obtained as \cite{bevington2002}
\begin{equation}
\Delta \nu =\left[ \left. \frac{1}{2} \frac{d^2 \chi^2(\nu)}{d\nu^2} \right|_{\nu_\mathrm{best}} \right]^{-1/2}.
\end{equation}
This yields  $\Delta\nu \simeq 0.2$, in good agreement with the bootstrap error estimation. We keep this value as the final estimation of the uncertainty of $\nu$.
\quad We have verified that extending the scaling function to third order in the scaling variable $u$, introducing an irrelevant scaling variable following Ref.~\cite{slevin_ohtsuki_crit_AL}, or allowing for nonlinear dependence of $\rho_c$ on $\omega$ do not significantly improve the fit quality: $\chi^2$ statistics (\ref{eq:residuals}) remains essentially unchanged even though the number of free parameters increases.

\quad It should be noted that although the near-field coupling that drives the delocalization transition depends on the spacing $d$ between the mirrors of the waveguide only weakly, the crossover distance below which it becomes dominant may depend on $d$. As a consequence, varying $d$ is not expected to suppress the transition, but may shift the critical density $\rho_c$. Therefore, the latter may not be universal. In contrast, we expect the critical exponent $\nu$ to remain unchanged as far as the condition $k_0 d < \pi$ is fulfilled. 

\section{Anderson localization at low atomic number density}\label{app:2dsc_tm}

\begin{figure}
    \centering
    \includegraphics[width=1\columnwidth]{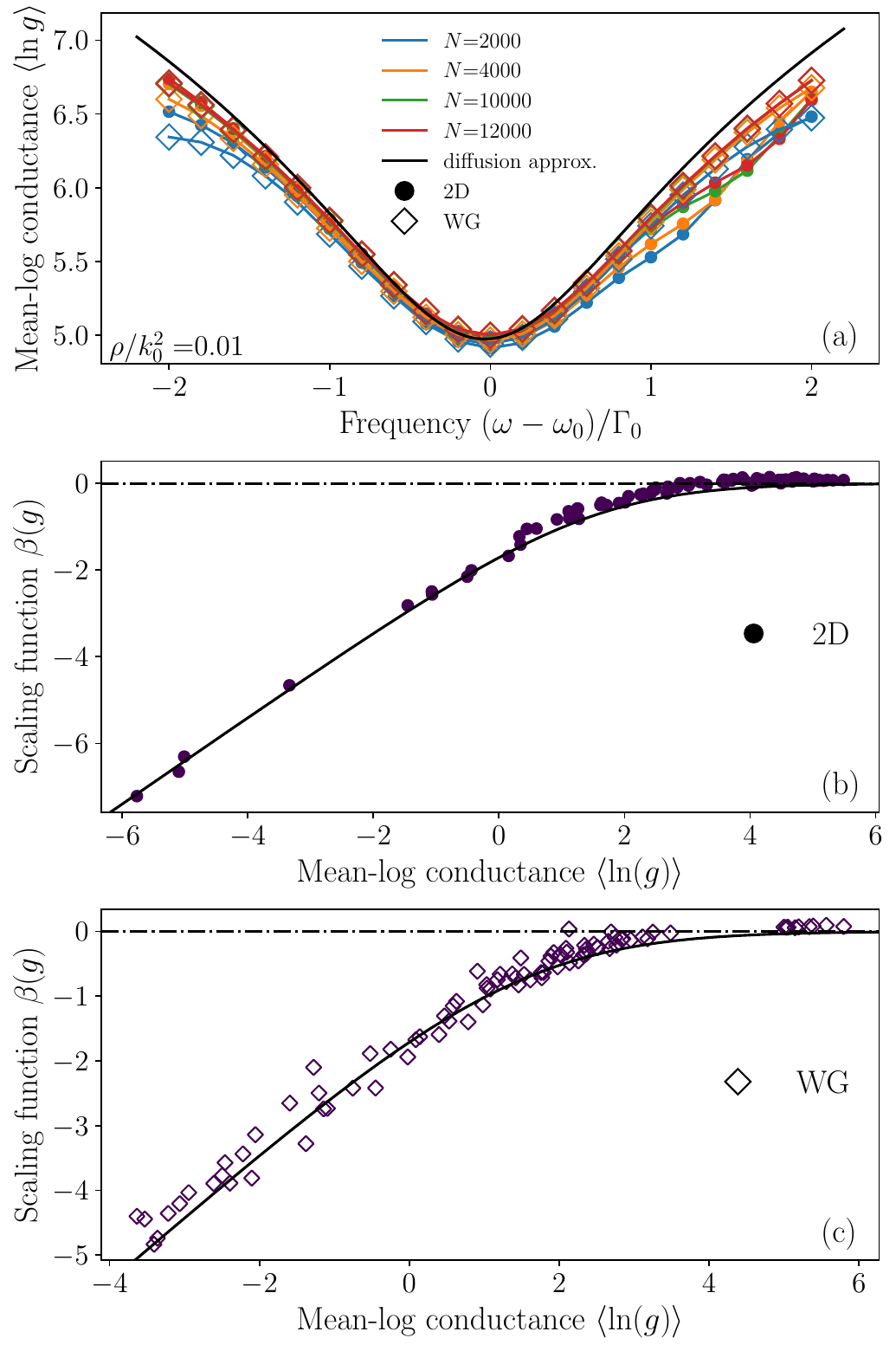}
    \caption{(a) Average logarithm of conductance $\langle \ln g \rangle$ as a function of frequency for different system sizes, for both WG (open diamonds) and 2D (dots) modes at low density $\rho/k_0^2=0.01$. The black line corresponds to the conductance computed in the diffusion approximation, using the independent scattering approximation (ISA, see Sec.~\ref{app:univ_cond}). (b) Scaling function $\beta(g)=d\langle \ln g \rangle/d\ln k_0 R$ for 2D  modes as a function of $\langle \ln g \rangle$ computed by a finite-difference approximation of the derivative at different small densities $\rho/k_0^2\leq0.5$ and considering only modes with frequencies satisfying $(\omega_n-\omega_0)/\Gamma_0\in[-0.75,1]$, \textit{i.e.} the modes least impacted by finite-size effects. The solid black curve corresponds to the analytical expectation from the scaling theory of Anderson localization in 2D:~$\beta(g)=1-(1+g)\ln (1 + g^{-1})$~\cite{MullerDelande2010} shifted horizontally to fit the numerical data. (c) Same as (b) but for WG modes.}\label{fig:2D_vs_TM_low_rhok}
\end{figure}

\begin{figure}
    \centering
    \includegraphics[width=1\columnwidth]{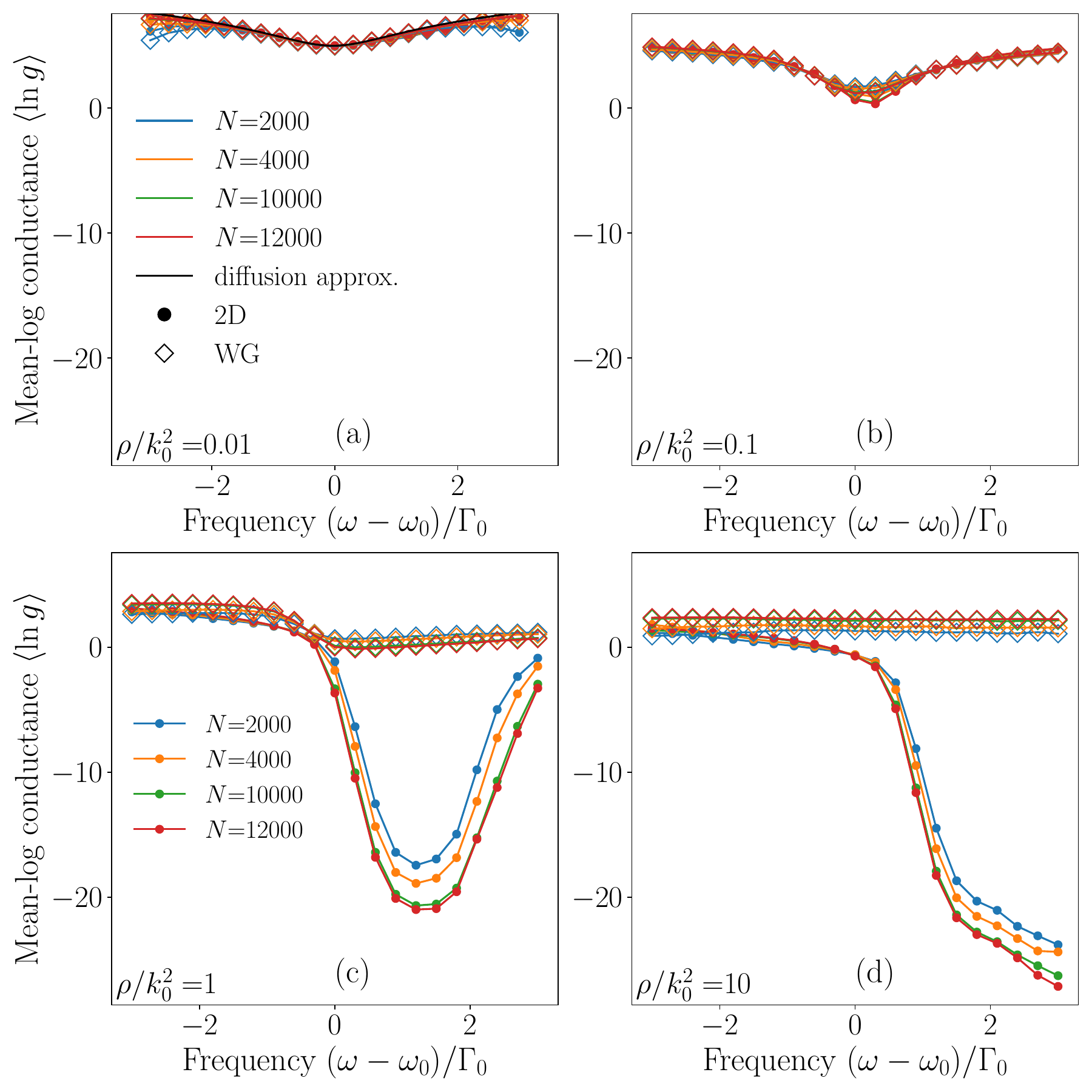}
    \caption{Average logarithm of conductance $\langle \ln g \rangle$ as a function of frequency for different system sizes, for both WG (open diamonds) and 2D (dots) modes at different densities $\rho/k_0^2=0.01$ (a), $0.1$ (b), $1$ (c), $10$ (d). }\label{fig:2D_vs_TM_conductance}
\end{figure}

\quad In this section, we demonstrate that both 2D and WG modes exhibit similar behavior and scaling expected for Anderson localization of scalar waves in 2D as long as the number density of atoms remains low. For each quasi-mode $|\Psi_n\rangle$, we define the Thouless conductance $g_n = \delta\omega_n/\Delta\omega_n$ with the mode spectral width $\delta\omega_n = \Gamma_n$ and the spacing between ordered mode frequencies $\Delta\omega_n = |\omega_{n+1} - \omega_{n}|$~\cite{thouless1977,Gang4,wang11,skipetrov_absence_2014}. As an example, we show $\langle \ln g \rangle$ at a low density $\rho/k_0^2 = 0.01$ in Fig.~\ref{fig:2D_vs_TM_low_rhok}(a), for both WG and 2D modes. In two dimensions, all states are in principle localized in the thermodynamic limit, but the localization length $\xi$ may exceed any practical system size $R$. Consequently, at low densities both systems exhibit the same quasi-extended behavior and the conductance agrees with the one computed in the diffusion approximation [see the solid black curve in Fig.~\ref{fig:2D_vs_TM_low_rhok}(a)], with the transport mean free path estimated within the independent scattering approximation (see Sec.~\ref{app:univ_cond}). In contrast, at higher densities $\rho/k_0^2 \gtrsim 1$, 2D results clearly exhibit a dip, with $\langle \ln g \rangle$ taking large negative values for $(\omega-\omega_0)/\Gamma_0 \gtrsim 1$ [see dots in Fig.~\ref{fig:2D_vs_TM_conductance} (c) and (d)], where the scattering is the strongest and $\xi$ becomes shorter than $R$, making Anderson localization effects important. Whereas such a behavior is expected in 2D, it is not at all followed by the Thouless conductance computed for WG modes. Instead, $\langle \ln g \rangle$ remains relatively large at all frequencies, signaling the suppression of Anderson localization at high atomic densities.

\quad The scaling function $\beta(g) = d \langle \ln g \rangle/d\ln k_0 R$ for both WG modes and 2D modes, obtained by a finite-difference approximation of the derivative, is shown in Fig.~\ref{fig:2D_vs_TM_low_rhok}(b) and (c) for low densities ($\rho/k_0^2<0.5$). As expected, both cases lead to qualitatively similar behavior with the numerical data for different $R$ and $\rho$ collapsing on a single curve, which, in its turn, matches the analytical expectation for Anderson localization in 2D~\cite{MullerDelande2010}.

\section{Finite-size effects in the computation of the scaling function}

\begin{figure}
    \centering
    \includegraphics[width=1\columnwidth]{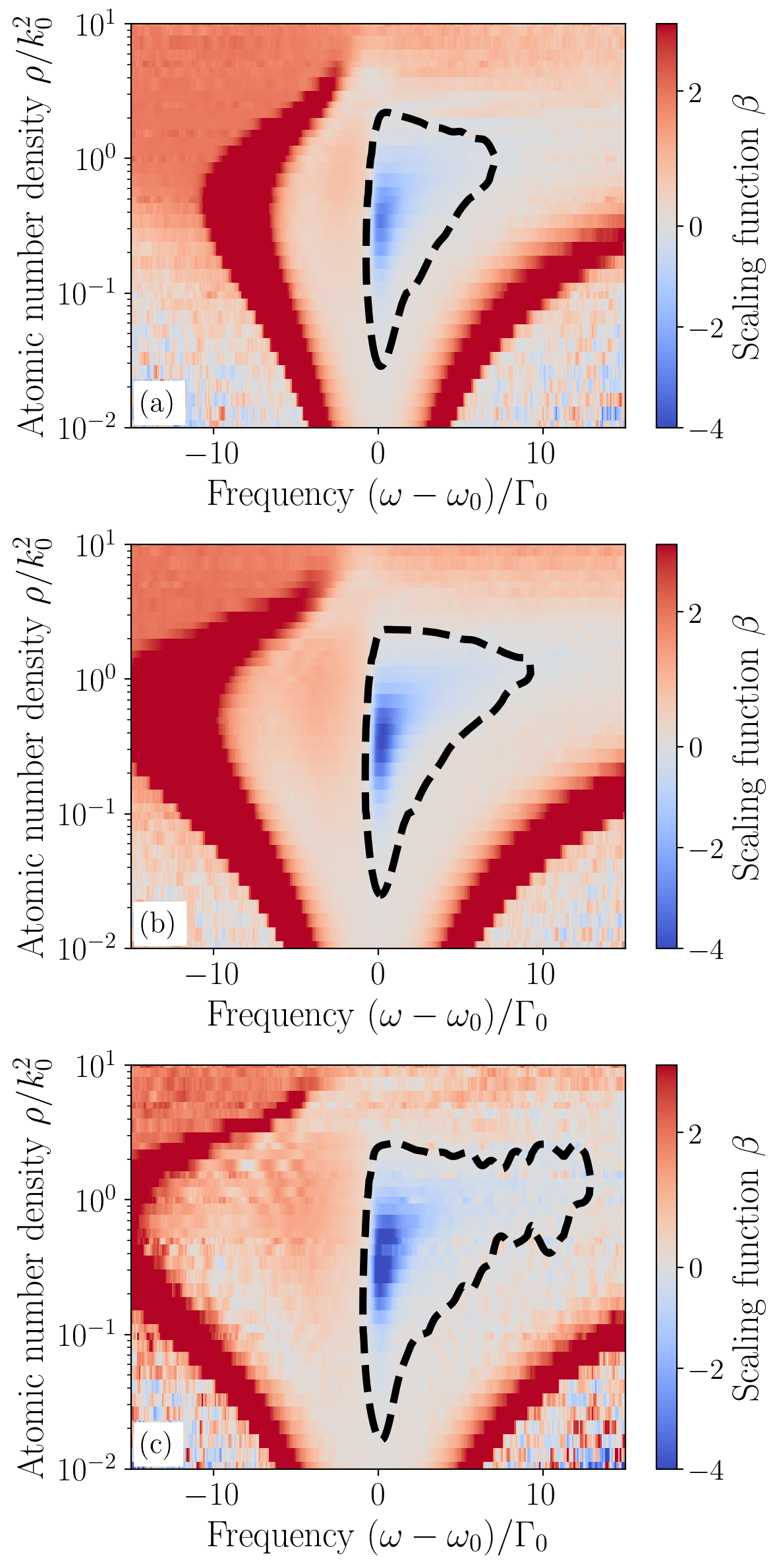}
    \caption{Scaling function $\beta$ for WG modes as a function of frequency $\omega$ and atomic number density $\rho$. $\beta$ is computed by a finite-difference approximation of the derivative for ensembles of $N = 4000$ and $2000$ atoms (a), $N = 10000$ and $4000$ atoms (b) or $N = 12000$ and $10000$ atoms (c). The dashed black line shows a contour $\beta=0$.}\label{fig:beta_func_finite_size}
\end{figure}

\quad The scaling function map $\beta(\omega,\rho)$ shown in Fig.~4(a) of the main text is subject to finite-size effects. Figure~\ref{fig:beta_func_finite_size} shows the same map obtained for different system sizes. The first observation is that the two red regions  corresponding to high values of $\beta \gtrsim 3$ shift to larger (in absolute value) frequencies as the system size increases, whereas the rest of the map remains qualitatively the same. These regions correspond to quasi-modes that are the most affected by finite-size effects and their shift away from the central, most interesting part of the map where $\beta < 0$, suggests that they should become irrelevant in the thermodynamic limit $N \to \infty$.
The second interesting observation is the fact that the upper part of the contour $\beta=0$ flattens as the system size increases. This may indicate that the critical density $\rho_c$ does not depend on frequency $\omega$ (\textit{i.e.}, $\delta \rho_c = 0$ in Sec.\ \ref{app:fitting}) in the thermodynamic limit, and that the weak but detectable dependence of $\rho_c$ on $\omega$ (\textit{i.e.}, $\delta \rho_c \ne 0$ in Sec.\ \ref{app:fitting}) is a finite-size effect.

\section{Conductance at low density}
\label{app:univ_cond}

\quad In this section we derive an expression for the conductance $g=\delta\omega/\Delta \omega$ in the limit of low density $\rho$ and in the diffusion approximation, and show that it is independent of the system size $R$. The modal decomposition of the solution of the diffusion equation for light intensity $I(\vec{r},t)$ in a disk of radius $R$ with boundary conditions $\left. I(\vec{r},t) \right|_{r = R} = 0$ yields (see, \textit{e.g.}, Ref.~\cite{brown2011fourier}, Sec.~9.93)
\begin{equation}
    \begin{aligned}
        I(\vec{r},t) = \sum_{m=1}^{\infty} f_{m}(\vec{r}) e^{-\alpha_{0m}^2D/R^2 t}
    \end{aligned}
\end{equation}
where $\alpha_{0m}$ is the $m$-th zero of the Bessel function of order $0$. At long times, the dynamics is dominated by the $m = 1$ term, which defines the typical spectral width of quasi-modes in the diffuse regime: $\delta\omega=\alpha_{01}^2 D / R^2$. We now recall that $D = v_E \ell/2$ with $v_E$ the energy transport velocity and $\ell$ the transport mean free path~\cite{lagendijk_resonant_1996}. For atomic scatterers, the latter is given by
\begin{equation}
    \begin{aligned}
        \ell &= \frac{1}{\rho \sigma_\mathrm{sc}}\\
    \end{aligned}
\end{equation}
where
\begin{equation}
    \begin{aligned}
        \sigma_\mathrm{sc} &= k_0 \mathrm{Im}[\alpha_\mathrm{at}]\\
        \alpha_\mathrm{at} &= -4 / k_0^ 2 (\Gamma_0 / 2) / (\delta + i\Gamma_0/2)\\
     \delta &= \omega-\omega_0   
    \end{aligned}
\end{equation}
are the scattering cross-section, the near-resonance atomic polarizability, and the frequency detuning from the atomic resonance, respectively. To estimate the average mode spacing  $\Delta\omega$, we notice that  $\Delta\omega={(\rho_\mathrm{DOS} \pi R^2)}^{-1}$,  where
$\rho_\mathrm{DOS} = k_0/2\pi v_E$ is the density of optical states in 2D.

\quad Combining the above results, we obtain
\begin{equation}
    \begin{aligned}
        g &= \frac{\delta \omega}{\Delta \omega}  =  \alpha_{01}^2 \pi D \rho_\mathrm{DOS} = \alpha_{01}^2 \frac{k_0\ell}{4} \\
          &= \frac{\alpha_{01}^2}{16 \rho/k_0^2}
           \left[1+{(2\delta /\Gamma_0)}^2\right]
    \end{aligned}
\end{equation}
We compare the logarithm of this expression with numerical results for $\langle \ln g \rangle$ in Fig.~\ref{fig:2D_vs_TM_low_rhok}(a),
multiplying it by an additional factor of about 4, which most likely results from the difference in treating fluctuations of $g$. Indeed, our numerical analysis yields a statistical average $\langle \ln g \rangle$ for $g$ fluctuating from one atomic configuration to another.  These fluctuations make $\langle \ln g \rangle$ different from  $\ln \langle g \rangle$ even when both averages are computed from the same numerical data. The calculation presented in this section is even less rigorous because the computed $g$ corresponds to some typical value different from both $\exp(\langle \ln g \rangle)$ and $\langle g \rangle$.

\section{IPR and fractal dimensions}

\quad To analyze the inverse participation ratios (IPR) and fractal dimensions $D_q$  of quasi-modes $|\Psi_n\rangle$, we define a set of $N_f$ equally spaced frequencies $\omega_k$, $k=1, \hdots, N_f$. For each $\omega_k$, system size $R_l$ (with $l=1, \hdots, N_R$), and each of $N_\mathrm{rea}$ independent atomic configurations, we introduce $\mathcal{J}_r(\omega_k)
= \left\{ j \in \{1,\dots,N\} : |\omega_j - \omega_k| < \tfrac{\Delta}{2} \right\}$, of cardinality $n_r = |\mathcal{J}_r(\omega_k)|$, 
where the modes are sorted by increasing frequencies $\omega_j$ and $\Delta = \Gamma_0/4$. We then compute the average of the logarithm of IPR as
\begin{equation}
    \langle \ln \mathrm{IPR}(q) \rangle (\omega_k ; R_l) = \frac{1}{N_\mathrm{rea}}\sum_{r=1}^{N_\mathrm{rea}}\frac{1}{n_r} \sum_{j\in\mathcal{J}_r(\omega_k)}  \ln \mathrm{IPR}_j(q)
\end{equation}
where $\mathrm{IPR}_j(q)=\sum_{a=1}^N|\Psi_j(\vec{r}_a)|^{2q}$, assuming normalized eigenvectors $|\Psi_j\rangle$. To extract the fractal dimension $D_q$, we perform a weighted linear fit of $\langle \ln{\mathrm{IPR}(q)}\rangle$ as a function of system size $R_l$ (with $l=1, \hdots, N_R$). Writing
\begin{equation}
\langle \ln{\mathrm{IPR}(q)}\rangle (\omega_k ; R_l) = a(\omega_k) \ln k_0R_l + b(\omega_k),
\end{equation}
the fit is obtained by minimizing a weighted least-squares cost function with weights $w_{k,l} =1/\sigma^2_{k,l}$, where $\sigma^2_{k,l}$ is the variance of $\langle \ln \mathrm{IPR}(q) \rangle(\omega_k ; R_l)$, computed in the same way as the variance of the logarithm of conductance in Sec.~\ref{app:error}. 

\quad The slope $a(\omega_k)$ is estimated from the weighted covariance
\begin{equation}
    a(\omega_k) =
    \frac{
    \sum_l w_{k,l} (x_l - \bar{x})(y_{k,l} - \bar{y}_k)
    }{
    \sum_l w_{k,l} {(x_l - \bar{x})}^2
    },
\end{equation}
where $x_l = \ln k_0R_l$, $y_{k,l} = \langle \ln \mathrm{IPR}(q) \rangle(\omega_k; R_l)$, and $\bar{x}$, $\bar{y}_k$ are the corresponding weighted means. The variance of the slope follows directly from the inverse weighted variance of $x$:
\begin{equation}
\sigma_a^2(\omega_k) =
    {\left[
    {\sum_l w_{k,l} (x_l - \bar{x})}^2
    \right]}^{-1}.
\end{equation}

The fractal dimension is then given by

\begin{equation}
    D_q(\omega_k) = \frac{a(\omega_k)}{1 - q},
\end{equation}
and its uncertainty is propagated linearly as
\begin{equation}
    \Delta D_q(\omega_k) = \frac{\sigma_a(\omega_k)}{|1 - q|}.
\end{equation}

\begin{figure}
    \centering
    \includegraphics[width=1.0\columnwidth]{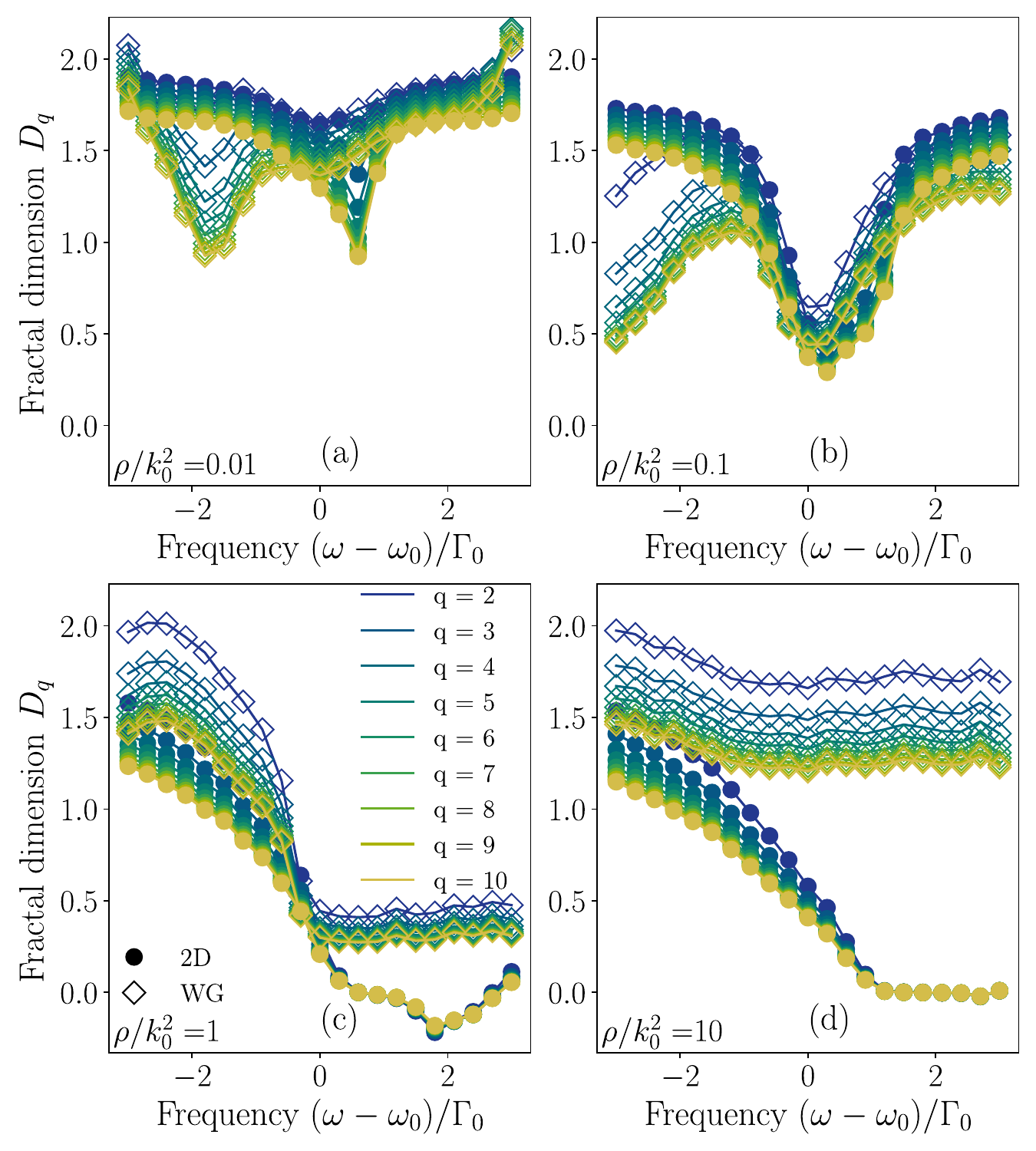}
    \caption{Fractal dimensions $D_q$ as functions of frequency for different atom numbers $N$ and orders $q\ge 2$, for both WG (open diamonds) and 2D (dots) modes, at density $\rho/k_0^2 =$ 0.01 (a), 0.1 (b), 1 (c) and 10 (d). For readability, the error-bars are not displayed.
    }\label{fig:Dq_tot}
\end{figure}

Figure~\ref{fig:Dq_tot} shows that for $q > 2$, $D_q$ exhibits qualitatively the same behavior as for $q = 2$. In particular, a clear difference emerges between WG and 2D modes at high density $\rho/k_0^2 > 1$.

\section{Typical mode profiles}

\begin{figure*}
    \centering
    \includegraphics[width=0.75\textwidth]{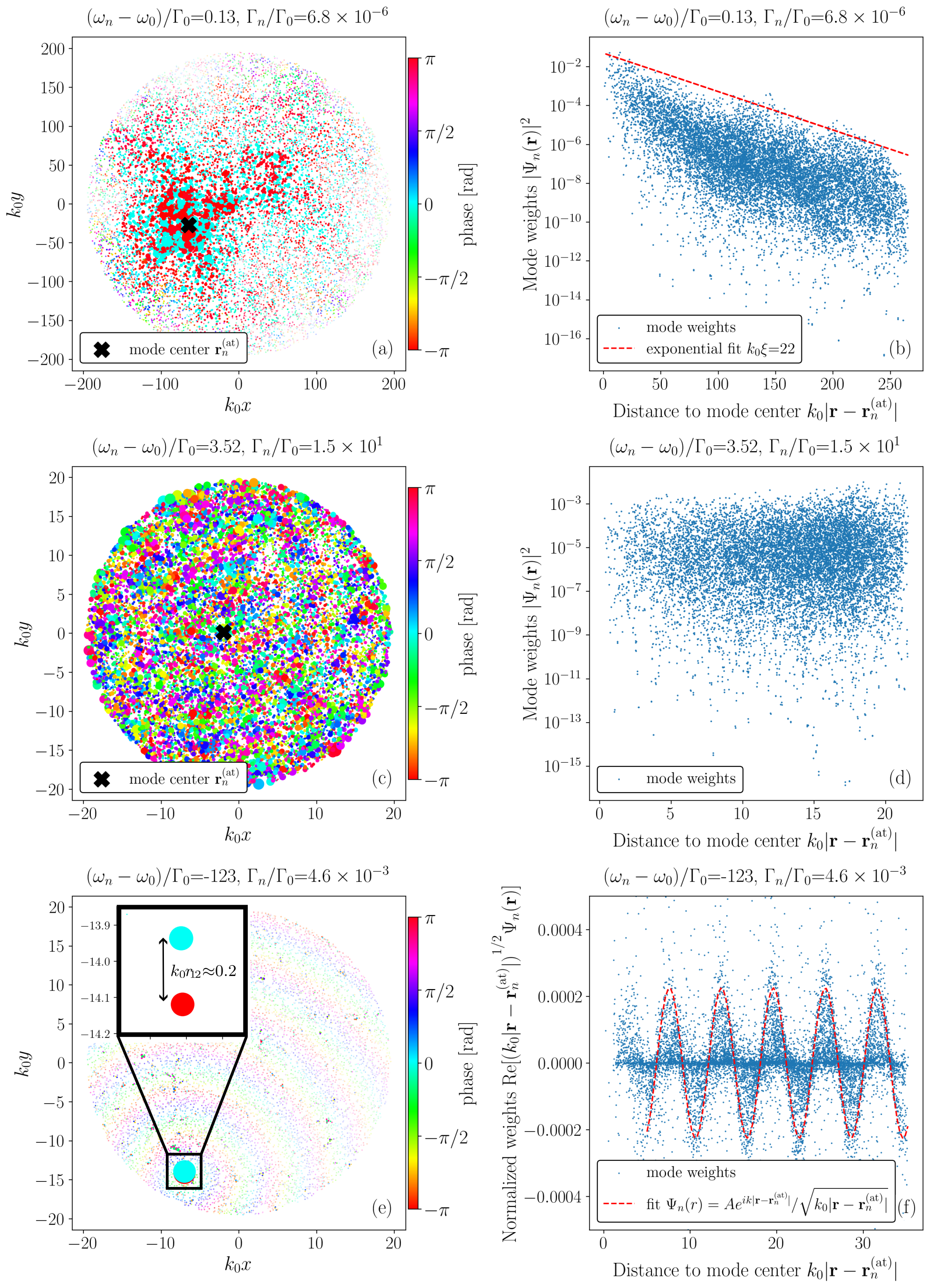}
    \caption{
    (Left panels a,c,e) Typical quasi-mode profiles. Each dot corresponds to an atomic position $\vec{r}_a$, its radius is proportional to the absolute value of mode amplitude on this atom $|\Psi_n(\vec{r}_a)|$, while the color encodes the phase of $\Psi_n(\vec{r}_a)$. The black cross marks the ``center of mass'' of the mode defined as $\vec{r}_n^{\text{(at)}}=\sum_a \vec{r}_a |\Psi_n(\vec{r}_a)|^2$.
    (Right panels b,d,f) Mode weight $|\Psi_n(\vec{r})|^2$ ($\mathrm{Re}[(\vec{r}-\vec{r}_n^{\text{(at)}})^{1/2} \Psi_n(\vec{r})]$ in (f))  as a function of distance to the mode center $\vec{r}_n^{\text{(at)}}$.
    (a,b) Anderson localized mode [$(\omega_n-\omega_0)/\Gamma_0=0.13$, $\Gamma_n/\Gamma_0=6.8\times10^{-6}$] obtained for $N=12000$ and $\rho/k_0^2=0.1$.
    (c,d) Extended mode [$(\omega_n-\omega_0)/\Gamma_0=3.52$, $\Gamma_n/\Gamma_0=15$] at $\rho/k_0^2=10$.
    (e,f) A subradiant mode spatially localized on a pair of closely located atoms, red and cyan dots in (e), [$(\omega_n-\omega_0)/\Gamma_0=-123$, $\Gamma_n/\Gamma_0=4.6 \times 10^{-3}$] at $\rho/k_0^2=10$. In (f) the weights are rescaled to remove the geometrical decay of the mode amplitude $|\Psi_n(\vec{r})| \propto (\vec{r}-\vec{r}_n^{\text{(at)}})^{-1/2}$. The dashed line shows a cylindrical wave with an effective wavenumber $k=1.05k_0$ adjusted to obtain the best agreement with the numerical data.
}\label{fig:mode_profiles}
\end{figure*}

\quad Figure~\ref{fig:mode_profiles} displays spatial profiles of typical WG modes at both low and high densities of atoms, corroborating the conclusions of the main text. In particular, exponentially localized states are present due to the phenomenon of Anderson localization at low density [Fig.~\ref{fig:mode_profiles}(a, b)], whereas at a high density, most of the modes are extended [Fig.~\ref{fig:mode_profiles}(c, d)], even though some subradiant quasi-modes localized on pairs of closely located atoms always exist [Fig.~\ref{fig:mode_profiles}(e, f)].
The spatial profiles of the latter resemble cylindrical waves emanating from the pair location. These quasi-modes do not show any significant dependence on sample size and do not influence the scaling analysis presented in the main text.

\quad It is interesting to note that mode amplitudes $\Psi_n(\vec{r})$ of  the exponentially localized mode shown in  Fig.~\ref{fig:mode_profiles}(a,b) tend to be real (the phase mainly takes values close to 0 or $\pm \pi$). This corresponds to a small complexity of the mode and represents an additional signature of Anderson localization~\cite{modal_complexity}. In contrast, extended modes [Fig.~\ref{fig:mode_profiles}(c,d)] and modes localized on pairs of atoms [Fig.~\ref{fig:mode_profiles}(e,f)] have complex amplitudes $\Psi_n(\vec{r})$ with phases more or less uniformly distributed between $-\pi$ and $\pi$, which corresponds to a large complexity, as expected for delocalized modes~\cite{modal_complexity}. 

\section{Correspondence between the localization of atomic quasi-modes and localization of the electromagnetic field}
 
\begin{figure*}
    \centering
    \includegraphics[width=1\textwidth]{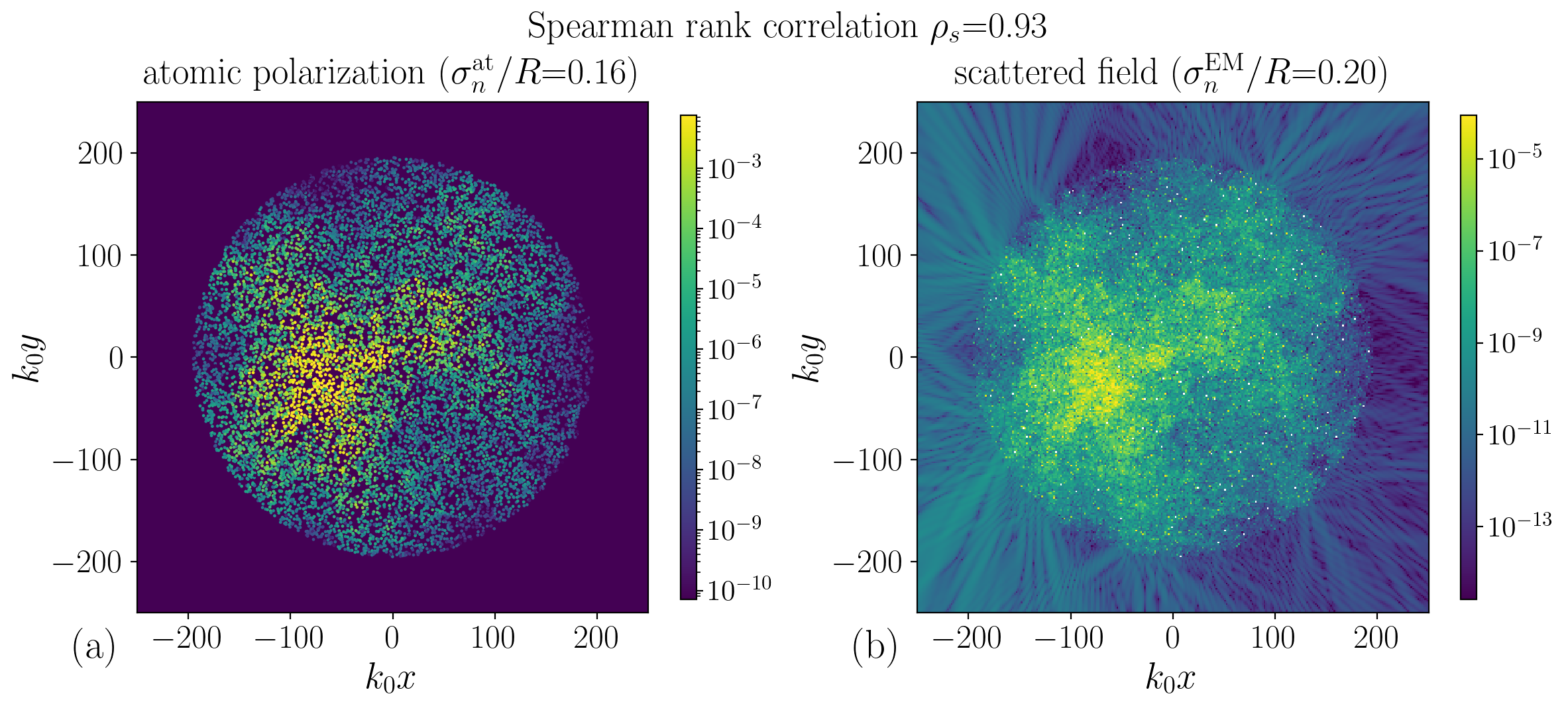}
    \caption{
    (a) Atomic polarization profile $|\Psi_n|^2$ of an Anderson localized mode [$(\omega_n-\omega_0)/\Gamma_0=0.13$, $\Gamma_n/\Gamma_0=6.8\times10^{-6}$] obtained for $N=12000$ and $\rho/k_0^2=0.1$.
    (b) Intensity profile $|E_n|^2$ of the field radiated by $\Psi_n$.
    The correlation between the two spatial profiles is quantified by Spearman's rank correlation coefficient $\rho_s$. Their spatial localization is measured by $\sigma_n$ defined by Eqs.~\eqref{eq1}--\eqref{eq4}. 
    }\label{fig:profil_loc}
\end{figure*}

\begin{figure*}
    \centering
    \includegraphics[width=1\textwidth]{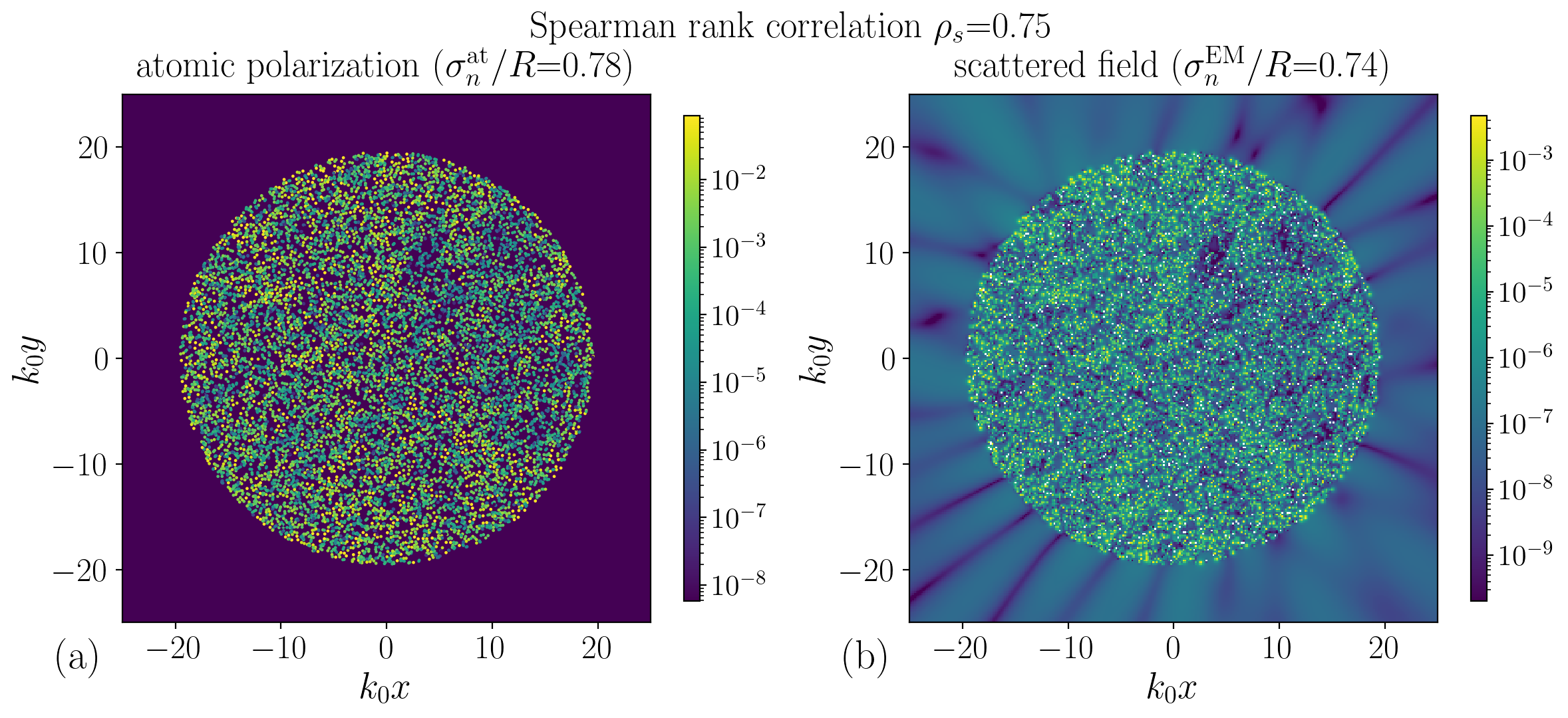}
    \caption{
    Same as Fig.\ \ref{fig:profil_loc} but for an extended mode [$(\omega_n-\omega_0)/\Gamma_0=3.5$, $\Gamma_n/\Gamma_0=15$] at $\rho/k_0^2=10$.
    }\label{fig:profil_deloc}
\end{figure*}

\begin{figure*}
    \centering
    \includegraphics[width=1\textwidth]{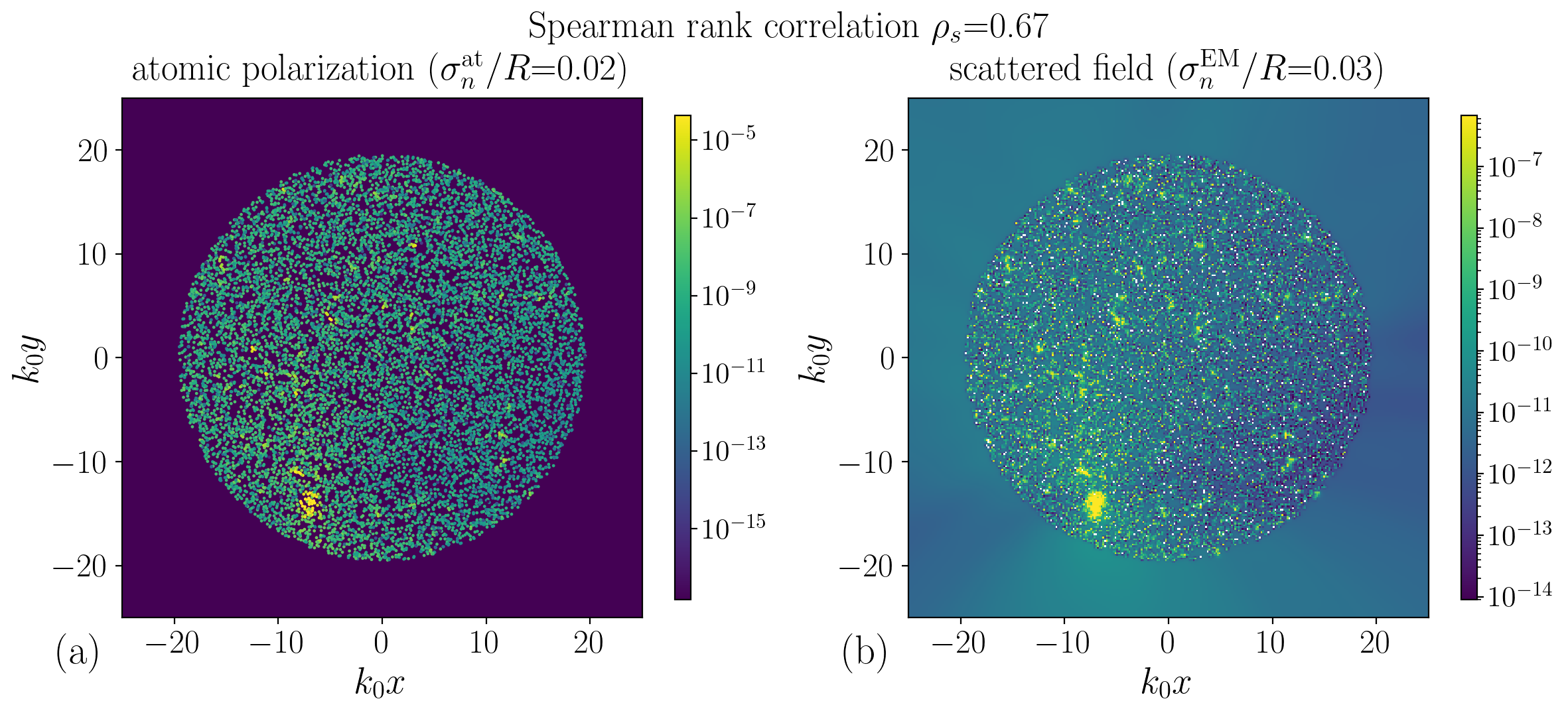}
    \caption{
    Same as Figs.\ \ref{fig:profil_loc} and \ref{fig:profil_deloc} but for a mode localized on a pair of closely located atoms [$(\omega_n-\omega_0)/\Gamma_0=-123$, $\Gamma_n/\Gamma_0=4.6 \times 10^{-3}$] at $\rho/k_0^2=10$.
    }\label{fig:profil_pair}
\end{figure*}

\quad The electric field $E_n$ radiated by an eigenmode $\Psi_n$ is
\begin{equation}
    E_n (\mathbf{r}) = \frac{\hbar \Gamma_0}{2 \mu_d} \sum_{a = 1}^{N} \tilde{\mathcal{G}}^{(\mathrm{WG})} (\mathbf{r} -   \mathbf{r}_a) \Psi_n (\mathbf{r}_a)
   \label{en}
\end{equation}
where $\mu_d$ is the dipole moment element of the atomic transition. Thus, $|E_n(\vec{r})|^2$ is a linear combination of $\Psi_n (\vec{r}_a) \Psi_n^* (\vec{r}_b)$ and not only a function of $|\Psi_n (\vec{r}_a)|^2$. For this reason, the standard correlation coefficients do not correctly reflect the link between $|E_n(\vec{r})|^2$ and $|\Psi_n(\vec{r})|^2$. Instead, we propose to use Spearman's rank correlation which measures the strength of a monotonic association between two variables. Spearman's rank correlation coefficient close to $\pm 1$ indicates a strong increasing/decreasing monotonic relationship between two variables, while statistically independent variables lead to 0. We compare the modes displayed in Fig.\ \ref{fig:mode_profiles} and the corresponding intensity maps computed using Eq.\ (\ref{en}) in Figs.\ \ref{fig:profil_loc}--\ref{fig:profil_pair}. The comparison reveals qualitative agreement between spatial profiles of $|\Psi_n |^2$ and $|E_n|^2$ with Spearman's rank correlation coefficient superior to $0.67$ for all modes (Anderson localized mode in Fig.\ \ref{fig:profil_loc}, extended mode in Fig.\ \ref{fig:profil_deloc}, mode localized on a pair of atoms in Fig.\ \ref{fig:profil_pair}), indicating a strong monotonic association between $|\Psi_n |^2$ and $|E_n|^2$.\\

\quad Spatial localization of both the atomic polarization and the corresponding electromagnetic field can be characterized by their mean-square extents
\begin{align}
(\sigma_n^{(\mathrm{at})})^2 &= \sum_{a = 1}^{N} | \mathbf{r}_a   - \mathbf{r}_n^{(\mathrm{at})} |^2 | \Psi_n (\mathbf{r}_a) |^2   \label{eq1}
  \\
 \mathbf{r}_n^{(\mathrm{at})} &= \sum_{a = 1}^{N} \mathbf{r}_a |
  \Psi_n (\mathbf{r}_a) |^2
  \label{eq2}
\end{align}
\begin{align}
(\sigma_n^{(\mathrm{EM})})^2
    &= \int_{| \mathbf{r} | \leqslant R} d^2
  \mathbf{r} | \mathbf{r} - \mathbf{r}_n^{(\mathrm{EM})} |^2 | E_n (\mathbf{r})
  |^2
  \\
  &\times \left(\int_{| \mathbf{r} | \leqslant R} d^2 \mathbf{r} | E_n (\mathbf{r})
  |^2 \right)^{-1}
  \label{eq3}
  \\
  \mathbf{r}_n^{(\mathrm{EM})} &= \int_{| \mathbf{r} | \leqslant R} d^2
  \mathbf{r}\,  \mathbf{r} | E_n (\mathbf{r}) |^2 / \int_{| \mathbf{r} |
  \leqslant R} d^2 \mathbf{r} | E_n (\mathbf{r}) |^2
  \label{eq4}
\end{align}
 
Both $\sigma_n^{(\mathrm{at})}$ and $\sigma_n^{(\mathrm{EM})}$ are
well-defined and finite: the sums in Eqs.\ (\ref{eq1}) and (\ref{eq2}) assume normalized eigenvectors
$\sum_{a = 1}^{N} | \Psi_n (\mathbf{r}_a) |^2 = 1$, while the integrals in Eqs.\ (\ref{eq3}) and (\ref{eq4}) are restricted to the disk $|\mathbf{r}| \leq R$
containing the atoms. $\sigma_n^{(\mathrm{at})}$,  $\sigma_n^{(\mathrm{EM})} \ll R$ signals strongly localized modes, whereas $\sigma_n^{(\mathrm{at})}$,  $\sigma_n^{(\mathrm{EM})} \approx R$ for extended modes.

\begin{figure}

        \includegraphics[width=\linewidth]{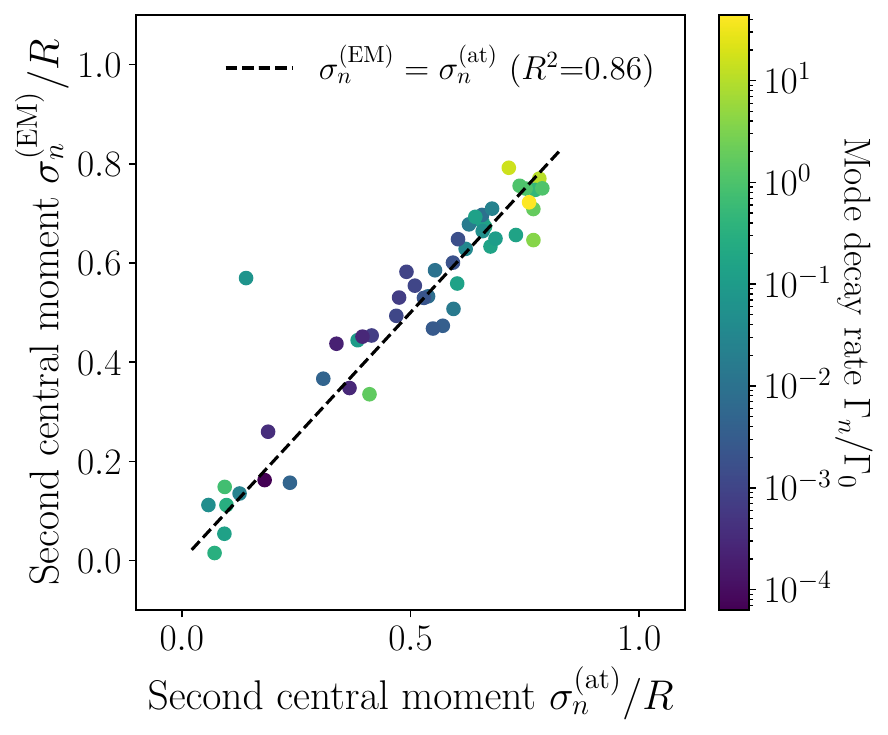}

        \includegraphics[width=\linewidth]{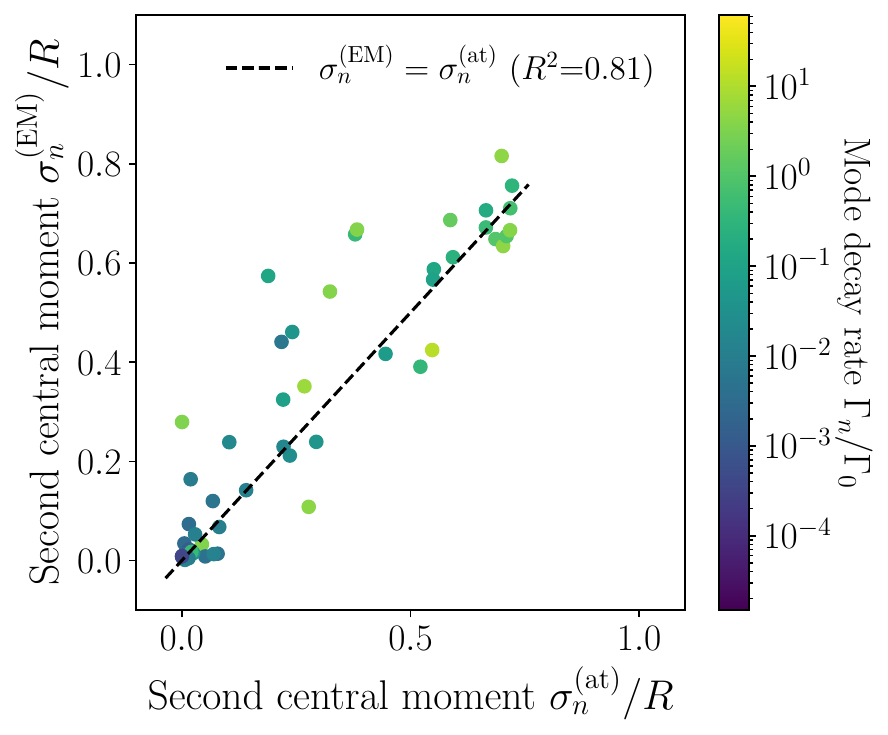}
    \caption{Root-mean-square spatial extent of the electromagnetic field (vertical axis) as a function of the root-mean-square spatial extent of the atomic polarization (horizontal axis) as defined by Eqs.~\eqref{eq1}--\eqref{eq4},
    computed for 50 randomly sampled modes in an ensemble of $N=12000$ atoms at $\rho / k_0^2 = 0.1$ [panel (a),  Anderson localization regime] and $\rho / k_0^2 = 10$ [panel (b), extended regime]. Both extended modes and modes localized on pairs of closely located atoms contribute in the latter case.
    The black dashed line corresponds to the relation $\sigma_n^{(\mathrm{EM})} = \sigma_n^{(\mathrm{at})}$ and describes the numerical data with a good regression coefficient $R^2 > 0.8$.
    Color scale corresponds to the decay rate of quasi-modes.
}\label{fig:extent}
\end{figure}

\quad
In order to demonstrate that the spatial localization of electromagnetic field is correlated with the spatial localization of atomic polarization, we compute their spatial extents defined by Eqs.\ \eqref{eq1}--\eqref{eq4} for a number of randomly selected modes at two different densities:  $\rho / k_0^2 = 0.1$ (Anderson localization regime) and $\rho / k_0^2 = 10$ (delocalized regime). Figure \ref{fig:extent} illustrates that a relation
$\sigma_n^{(\mathrm{EM})} \simeq \sigma_n^{(\mathrm{at})}$ describes the data with a
very good regression coefficient $R^2 > 0.8$, clearly demonstrating
a strong correlation between $\sigma_n^{(\mathrm{at})}$ and $\sigma_n^{(\mathrm{EM})}$.

\bibliography{bibliography}

\begin{thebibliography}{51}
\bibitem [{not()}]{note_eff_hamiltonian}%
  \BibitemOpen
  \href@noop {} {}\bibinfo {note} {The effective Hamiltonian considered in this
  work is complex symmetric ($H_{\mathrm{eff}} = H_{\mathrm{eff}}^T$), which
  follows directly from the reciprocity of the Green's function. This implies
  that the left and right eigenvalues are identical and that left eigenvectors
  are not independent but are related to right eigenvectors by complex
  conjugation (up to normalization conventions). Consequently, the study of
  right eigenvalues and right eigenvectors is sufficient.}\BibitemShut {Stop}%
\end{thebibliography}

\end{document}